%% file: main.tex
\pdfoutput=1
\documentclass{article}

\usepackage{PRIMEarxiv}

\usepackage[utf8]{inputenc}
\usepackage[T1]{fontenc}
\usepackage{amsmath}
\usepackage{amsthm}
\usepackage{booktabs}
\usepackage{tabularx}
\usepackage{float}
\usepackage{microtype}
\usepackage{graphicx}
\usepackage{xspace}
\xspaceaddexceptions{\%}
\usepackage[round]{natbib}
\usepackage{xurl}
\usepackage{hyperref}
\hypersetup{colorlinks=true, linkcolor=blue, citecolor=blue, urlcolor=blue}
\hypersetup{pdftitle={Ask More, Hear Less: How a Review Prompt Changes What Online Ratings Show},
  pdfauthor={Zihan Chen, Lei Nico Zheng, Di Zhu},
  pdfsubject={Online reviews, review solicitation and the informativeness of online ratings: evidence from Steam},
  pdfkeywords={online reviews, review solicitation, rating inflation, signal detection theory, electronic word of mouth, reputation systems, platform design, video games}}
\graphicspath{{figures/}}

\newtheorem{proposition}{Proposition}
\input{generated/numbers}

\title{Ask More, Hear Less:\\How a Review Prompt Changes\\What Online Ratings Show}

\author{
  Zihan Chen \\
  Stevens Institute of Technology \\
  Hoboken, NJ, USA \\
  \texttt{zchen61@stevens.edu} \\
  \And
  Lei Nico Zheng \\
  University of Massachusetts Boston \\
  Boston, MA, USA \\
  \texttt{lei.zheng@umb.edu} \\
  \And
  Di Zhu \\
  Stevens Institute of Technology \\
  Hoboken, NJ, USA \\
  \texttt{dzhu1@stevens.edu} \\
}

\begin{document}
\maketitle

\input{sections/00_abstract}
\input{sections/01_intro}
\input{sections/02_literature}
\input{sections/03_theory}
\input{sections/04_setting}
\input{sections/05_strategy}
\input{sections/06_results}

\input{sections/07_discussion}
\input{sections/08_limitations}
\input{sections/09_conclusion}

\bibliographystyle{plainnat}
\bibliography{references}

\clearpage
\appendix
\input{sections/10_appendix}

\end{document}

%% file: generated/numbers.tex
\newcommand{\AUCDid}{+0.02\xspace}
\newcommand{\AUCHi}{0.17\xspace}
\newcommand{\AUCLo}{$-$0.13\xspace}
\newcommand{\AUCaPost}{0.62\xspace}
\newcommand{\AUCaPre}{0.69\xspace}
\newcommand{\AUCbPost}{0.62\xspace}
\newcommand{\AUCbPre}{0.71\xspace}
\newcommand{\AddA}{81\xspace}
\newcommand{\AddB}{57\xspace}
\newcommand{\AgeDiff}{$-$0.25\xspace}
\newcommand{\AgeSE}{0.12\xspace}
\newcommand{\AgreeCost}{1.7\xspace}
\newcommand{\BigTTen}{$-$40\xspace}
\newcommand{\BigTTenHi}{4\xspace}
\newcommand{\BigTTenLo}{$-$73\xspace}
\newcommand{\BothCIHi}{$-$0.10\xspace}
\newcommand{\BothCILo}{$-$0.51\xspace}
\newcommand{\BothDiff}{$-$0.26\xspace}
\newcommand{\BothG}{143\xspace}
\newcommand{\BothLo}{$-$0.17\xspace}
\newcommand{\BothLoHi}{0.11\xspace}
\newcommand{\BothLoLo}{$-$0.62\xspace}
\newcommand{\BoundPost}{4.9\xspace}
\newcommand{\BoundPre}{6.3\xspace}
\newcommand{\BoundTwoPost}{14.2\xspace}
\newcommand{\BoundTwoPre}{26.3\xspace}
\newcommand{\CRDid}{+0.37\xspace}
\newcommand{\CRHi}{0.93\xspace}
\newcommand{\CRLo}{0.00\xspace}
\newcommand{\CRaPost}{1.25\xspace}
\newcommand{\CRaPre}{0.64\xspace}
\newcommand{\CRbPost}{0.74\xspace}
\newcommand{\CRbPre}{0.50\xspace}
\newcommand{\CWBelow}{5.8\xspace}
\newcommand{\CWExtra}{1.6\xspace}
\newcommand{\CWLongAdded}{11.6\xspace}
\newcommand{\CWLongBelow}{18.7\xspace}
\newcommand{\CWShortAdded}{71\xspace}
\newcommand{\CWShortBelow}{45\xspace}
\newcommand{\CWStd}{2.4\xspace}
\newcommand{\CarryW}{37\xspace}
\newcommand{\CarryWHi}{52\xspace}
\newcommand{\CarryWLo}{19\xspace}
\newcommand{\CfCrDiff}{+0.91\xspace}
\newcommand{\CfCrHi}{1.44\xspace}
\newcommand{\CfCrLo}{0.52\xspace}
\newcommand{\CfDpDiff}{+0.24\xspace}
\newcommand{\CfDpHi}{1.14\xspace}
\newcommand{\CfDpLo}{$-$0.78\xspace}
\newcommand{\CfFaEst}{$-$17\xspace}
\newcommand{\CfFaHi}{$-$11\xspace}
\newcommand{\CfFaLo}{$-$23\xspace}
\newcommand{\CfFixAbs}{4.9\xspace}
\newcommand{\CfFixFa}{20\xspace}
\newcommand{\CfFixHit}{30\xspace}
\newcommand{\CfHitEst}{$-$21\xspace}
\newcommand{\CfHitHi}{$-$2\xspace}
\newcommand{\CfHitLo}{$-$42\xspace}
\newcommand{\CfPropAbs}{1.9\xspace}
\newcommand{\CfPropCrDiff}{+0.32\xspace}
\newcommand{\CfPropCrHi}{0.83\xspace}
\newcommand{\CfPropCrLo}{$-$0.14\xspace}
\newcommand{\CfPropDpDiff}{+0.20\xspace}
\newcommand{\CfPropDpHi}{1.14\xspace}
\newcommand{\CfPropDpLo}{$-$0.80\xspace}
\newcommand{\CfPropFa}{7\xspace}
\newcommand{\CfPropFaEst}{$-$4\xspace}
\newcommand{\CfPropFaHi}{1\xspace}
\newcommand{\CfPropFaLo}{$-$9\xspace}
\newcommand{\CfPropHit}{13\xspace}
\newcommand{\CfPropHitEst}{$-$4\xspace}
\newcommand{\CfPropHitHi}{10\xspace}
\newcommand{\CfPropHitLo}{$-$21\xspace}
\newcommand{\CivPost}{1.3\xspace}
\newcommand{\CivPre}{23.6\xspace}
\newcommand{\CorrFixDid}{$-$0.09\xspace}
\newcommand{\CorrFixDidAbs}{0.09\xspace}
\newcommand{\CorrFixHi}{0.00\xspace}
\newcommand{\CorrFixLo}{$-$0.19\xspace}
\newcommand{\CorrFixPctDid}{$-$0.11\xspace}
\newcommand{\CorrFixPctHi}{$-$0.01\xspace}
\newcommand{\CorrFixPctLo}{$-$0.22\xspace}
\newcommand{\CorrFixaPost}{0.08\xspace}
\newcommand{\CorrFixaPre}{0.27\xspace}
\newcommand{\CorrFixbPost}{0.18\xspace}
\newcommand{\CorrFixbPre}{0.28\xspace}
\newcommand{\CorrFortyDid}{$-$0.13\xspace}
\newcommand{\CorrFortyHi}{$-$0.01\xspace}
\newcommand{\CorrFortyLo}{$-$0.28\xspace}
\newcommand{\CorrNatDid}{$-$0.03\xspace}
\newcommand{\CorrNatDidAbs}{0.03\xspace}
\newcommand{\CorrNatHi}{0.06\xspace}
\newcommand{\CorrNatLo}{$-$0.13\xspace}
\newcommand{\CorrNatLoAbs}{0.13\xspace}
\newcommand{\CorrNatN}{3,119\xspace}
\newcommand{\CorrNataPost}{0.22\xspace}
\newcommand{\CorrNataPre}{0.36\xspace}
\newcommand{\CorrNatbPost}{0.30\xspace}
\newcommand{\CorrNatbPre}{0.40\xspace}
\newcommand{\CutBeforeHi}{1.07\xspace}
\newcommand{\CutBeforeLo}{0.91\xspace}
\newcommand{\CutMain}{$-$0.36\xspace}
\newcommand{\CutOthHi}{+0.08\xspace}
\newcommand{\CutOthLo}{$-$0.01\xspace}
\newcommand{\CutTen}{1.33\xspace}
\newcommand{\CutThirty}{1.52\xspace}
\newcommand{\CutThou}{4.09\xspace}
\newcommand{\DADid}{+0.10\xspace}
\newcommand{\DAHi}{0.69\xspace}
\newcommand{\DALo}{$-$0.49\xspace}
\newcommand{\DAaPost}{0.44\xspace}
\newcommand{\DAaPre}{0.71\xspace}
\newcommand{\DAbPost}{0.43\xspace}
\newcommand{\DAbPre}{0.80\xspace}
\newcommand{\DFDid}{+0.8\xspace}
\newcommand{\DFaPost}{8\xspace}
\newcommand{\DFaPre}{14\xspace}
\newcommand{\DFbPost}{11\xspace}
\newcommand{\DFbPre}{18\xspace}
\newcommand{\DHDid}{$-$23\xspace}
\newcommand{\DHDidHi}{3\xspace}
\newcommand{\DHDidLo}{$-$50\xspace}
\newcommand{\DHaPost}{12\xspace}
\newcommand{\DHaPre}{42\xspace}
\newcommand{\DHbPost}{41\xspace}
\newcommand{\DHbPre}{47\xspace}
\newcommand{\DPDid}{$-$0.71\xspace}
\newcommand{\DPHi}{0.16\xspace}
\newcommand{\DPLo}{$-$1.89\xspace}
\newcommand{\DPaPost}{0.31\xspace}
\newcommand{\DPaPre}{0.86\xspace}
\newcommand{\DPbPost}{1.01\xspace}
\newcommand{\DPbPre}{0.86\xspace}
\newcommand{\DTnPost}{32\xspace}
\newcommand{\DTnPre}{72\xspace}
\newcommand{\DatePost}{March 2020\xspace}
\newcommand{\DatePostEnd}{January 2021\xspace}
\newcommand{\DatePreOne}{April 2017\xspace}
\newcommand{\DatePreOneEnd}{February 2019\xspace}
\newcommand{\DatePreTwo}{June 2019\xspace}
\newcommand{\DatePreTwoEnd}{September 2019\xspace}
\newcommand{\DecLo}{0.58\xspace}
\newcommand{\DecNoCeiling}{0.35\xspace}
\newcommand{\DecNoSensor}{0.32\xspace}
\newcommand{\DecObs}{0.26\xspace}
\newcommand{\DecPostObs}{0.20\xspace}
\newcommand{\DecPostPred}{0.19\xspace}
\newcommand{\DecPrePred}{0.60\xspace}
\newcommand{\DecProd}{0.31\xspace}
\newcommand{\DecRef}{1.01\xspace}
\newcommand{\DecTarget}{0.53\xspace}
\newcommand{\DetCountTen}{12\xspace}
\newcommand{\DetLogM}{4\xspace}
\newcommand{\DetPctM}{9\xspace}
\newcommand{\DetPctTen}{27\xspace}
\newcommand{\DetPctTwenty}{39\xspace}
\newcommand{\DetPreTen}{38\xspace}
\newcommand{\DetPreTwenty}{56\xspace}
\newcommand{\DetShareHi}{11\xspace}
\newcommand{\DetShareLo}{$-$30\xspace}
\newcommand{\DetShareTen}{19\xspace}
\newcommand{\EditPost}{4.2\xspace}
\newcommand{\EditPre}{13.3\xspace}
\newcommand{\EngDiff}{$-$0.28\xspace}
\newcommand{\EngF}{13\xspace}
\newcommand{\EngHi}{0.13\xspace}
\newcommand{\EngLo}{$-$0.74\xspace}
\newcommand{\EpG}{7\xspace}
\newcommand{\EpLess}{6\xspace}
\newcommand{\EqDiff}{$-$0.41\xspace}
\newcommand{\EqSE}{0.12\xspace}
\newcommand{\EtaImp}{0.55\xspace}
\newcommand{\EtaImpHi}{1.15\xspace}
\newcommand{\EtaImpLo}{0.18\xspace}
\newcommand{\EtaPair}{0.38\xspace}
\newcommand{\EtaPairHi}{0.83\xspace}
\newcommand{\EtaPairLo}{0.09\xspace}
\newcommand{\EvPostAbove}{0.09\xspace}
\newcommand{\EvPostBelow}{0.18\xspace}
\newcommand{\EvPostBelowDip}{0.07\xspace}
\newcommand{\EvPreAbove}{0.08\xspace}
\newcommand{\EvPreBelow}{0.06\xspace}
\newcommand{\ExtraPos}{96.4\xspace}
\newcommand{\FRGapDiff}{$-$0.48\xspace}
\newcommand{\FRGapDiffSE}{0.15\xspace}
\newcommand{\FaAllAUCPost}{0.64\xspace}
\newcommand{\FaAllAUCPre}{0.53\xspace}
\newcommand{\FaAllDid}{$-$1.8\xspace}
\newcommand{\FaAllHi}{0.9\xspace}
\newcommand{\FaAllLo}{$-$4.5\xspace}
\newcommand{\FaAllLoPost}{$-$0.20\xspace}
\newcommand{\FaAllLoPre}{$-$0.07\xspace}
\newcommand{\FaAllPostAbs}{2.9\xspace}
\newcommand{\FaAllPostSE}{0.8\xspace}
\newcommand{\FaAllPreAbs}{1.0\xspace}
\newcommand{\FaAllPreSE}{1.2\xspace}
\newcommand{\FaEpPost}{25\xspace}
\newcommand{\FaGapDid}{+0.9\xspace}
\newcommand{\FaGapHi}{7.6\xspace}
\newcommand{\FaGapLo}{$-$5.2\xspace}
\newcommand{\FaHPost}{$-$3.5\xspace}
\newcommand{\FaHPostSE}{1.4\xspace}
\newcommand{\FaMajPost}{7\xspace}
\newcommand{\FaNegPost}{+0.23\xspace}
\newcommand{\FaNegPre}{+0.25\xspace}
\newcommand{\FaPosPost}{+0.02\xspace}
\newcommand{\FaPosPre}{+0.17\xspace}
\newcommand{\FaSAUCPost}{0.43\xspace}
\newcommand{\FaSAUCPre}{0.50\xspace}
\newcommand{\FaSPost}{+2.6\xspace}
\newcommand{\FaSPostSE}{1.9\xspace}
\newcommand{\FaSPre}{+0.3\xspace}
\newcommand{\FaSPreSE}{1.9\xspace}
\newcommand{\FailCand}{727\xspace}
\newcommand{\FailDec}{18\xspace}
\newcommand{\FailEpisodes}{143\xspace}
\newcommand{\FailFOne}{176\xspace}
\newcommand{\FailFTwo}{87\xspace}
\newcommand{\FailGames}{63\xspace}
\newcommand{\FailMajor}{50\xspace}
\newcommand{\FailN}{464\xspace}
\newcommand{\FailTech}{125\xspace}
\newcommand{\FakeMax}{+0.14\xspace}
\newcommand{\FakeMin}{$-$0.15\xspace}
\newcommand{\FlowBothG}{143\xspace}
\newcommand{\FlowGames}{271\xspace}
\newcommand{\FlowLongT}{1.39\xspace}
\newcommand{\FlowLongU}{0.85\xspace}
\newcommand{\FlowNegX}{1.25\xspace}
\newcommand{\FlowPairPostG}{167\xspace}
\newcommand{\FlowPairPostGM}{1,241\xspace}
\newcommand{\FlowPairPreG}{222\xspace}
\newcommand{\FlowPairPreGM}{1,878\xspace}
\newcommand{\FlowPosX}{2.81\xspace}
\newcommand{\FlowTNeg}{1.57\xspace}
\newcommand{\FlowTPos}{4.17\xspace}
\newcommand{\FlowUNeg}{0.81\xspace}
\newcommand{\FlowUPos}{1.18\xspace}
\newcommand{\GainCon}{25\xspace}
\newcommand{\GainConHi}{45\xspace}
\newcommand{\GainConLo}{3\xspace}
\newcommand{\GainDid}{26\xspace}
\newcommand{\GainDidHi}{52\xspace}
\newcommand{\GainDidLo}{0\xspace}
\newcommand{\GainaPost}{12\xspace}
\newcommand{\GainaPre}{$-$3\xspace}
\newcommand{\GainbPostAbs}{12\xspace}
\newcommand{\GainbPre}{$-$1\xspace}
\newcommand{\GapDiff}{$-$0.40\xspace}
\newcommand{\GapDiffHi}{$-$0.21\xspace}
\newcommand{\GapDiffLo}{$-$0.57\xspace}
\newcommand{\GapDiffSE}{0.10\xspace}
\newcommand{\GapFPost}{82\xspace}
\newcommand{\GapFPre}{418\xspace}
\newcommand{\GapLoDiff}{$-$0.35\xspace}
\newcommand{\GapLoDiffHi}{$-$0.04\xspace}
\newcommand{\GapLoDiffLo}{$-$0.65\xspace}
\newcommand{\GapLoDiffSE}{0.15\xspace}
\newcommand{\GapPost}{$-$0.44\xspace}
\newcommand{\GapPre}{$-$0.04\xspace}
\newcommand{\GridMax}{99.4\xspace}
\newcommand{\GridMin}{94.2\xspace}
\newcommand{\HoldEarly}{180\xspace}
\newcommand{\HoldLate}{4.31\xspace}
\newcommand{\HoldLateMed}{0.95\xspace}
\newcommand{\HoldOther}{0.95\xspace}
\newcommand{\HoldShare}{100\xspace}
\newcommand{\HoldSplits}{40\xspace}
\newcommand{\HoldVal}{4.53\xspace}
\newcommand{\HoleDays}{15\xspace}
\newcommand{\HoleDiff}{$-$0.33\xspace}
\newcommand{\HoleGames}{3\xspace}
\newcommand{\HoleSE}{0.09\xspace}
\newcommand{\KwN}{340\xspace}
\newcommand{\KwPrec}{15\xspace}
\newcommand{\KwUn}{3\xspace}
\newcommand{\LOaPost}{0.57\xspace}
\newcommand{\LOaPre}{0.85\xspace}
\newcommand{\LObPost}{0.86\xspace}
\newcommand{\LObPre}{0.89\xspace}
\newcommand{\LOlgPost}{0.47\xspace}
\newcommand{\LOlgPre}{0.55\xspace}
\newcommand{\LOshPost}{0.51\xspace}
\newcommand{\LOshPre}{0.77\xspace}
\newcommand{\LabelDate}{November 1, 2026\xspace}
\newcommand{\LabelN}{3,264\xspace}
\newcommand{\LbCalmDpA}{0.4\xspace}
\newcommand{\LbCalmDpB}{0.6\xspace}
\newcommand{\LbCalmDpC}{0.2\xspace}
\newcommand{\LbCalmTierA}{13\xspace}
\newcommand{\LbCalmTierB}{10\xspace}
\newcommand{\LbCalmTierC}{3\xspace}
\newcommand{\LbFifDpA}{8.4\xspace}
\newcommand{\LbFifDpB}{10.1\xspace}
\newcommand{\LbFifDpC}{2.1\xspace}
\newcommand{\LbFifTierA}{34\xspace}
\newcommand{\LbFifTierB}{44\xspace}
\newcommand{\LbFifTierC}{17\xspace}
\newcommand{\LbTenDpA}{5.2\xspace}
\newcommand{\LbTenDpB}{7.2\xspace}
\newcommand{\LbTenDpC}{0.8\xspace}
\newcommand{\LbTenTierA}{29\xspace}
\newcommand{\LbTenTierB}{43\xspace}
\newcommand{\LbTenTierC}{8\xspace}
\newcommand{\LostHundred}{0.40\xspace}
\newcommand{\LostTwentyFive}{0.34\xspace}
\newcommand{\MainTTen}{+6\xspace}
\newcommand{\Margin}{0.05\xspace}
\newcommand{\McB}{0.42\xspace}
\newcommand{\McC}{0.41\xspace}
\newcommand{\McD}{0.40\xspace}
\newcommand{\McDiff}{$-$0.007\xspace}
\newcommand{\McDiffD}{$-$0.026\xspace}
\newcommand{\McDiffDHi}{$-$0.001\xspace}
\newcommand{\McDiffDLo}{$-$0.055\xspace}
\newcommand{\McDiffHi}{0.016\xspace}
\newcommand{\McDiffLo}{$-$0.034\xspace}
\newcommand{\McDiffLoNinety}{$-$0.029\xspace}
\newcommand{\McGames}{1,066\xspace}
\newcommand{\McLD}{0.35\xspace}
\newcommand{\McLDiff}{$-$0.032\xspace}
\newcommand{\McLDiffD}{$-$0.066\xspace}
\newcommand{\McLDiffDHi}{$-$0.030\xspace}
\newcommand{\McLDiffDLo}{$-$0.105\xspace}
\newcommand{\McLDiffHi}{0.000\xspace}
\newcommand{\McLDiffLo}{$-$0.065\xspace}
\newcommand{\McLDiffLoNinety}{$-$0.060\xspace}
\newcommand{\MinAfter}{11,163\xspace}
\newcommand{\MinAfterPre}{3,390\xspace}
\newcommand{\MinBefore}{2,316\xspace}
\newcommand{\MinBeforePre}{3,255\xspace}
\newcommand{\MixGM}{94.9\xspace}
\newcommand{\MixLL}{95.8\xspace}
\newcommand{\MixPre}{95.6\xspace}
\newcommand{\MixSixty}{96.9\xspace}
\newcommand{\MixTen}{96.0\xspace}
\newcommand{\MultCorrScore}{$-$0.34\xspace}
\newcommand{\MultG}{161\xspace}
\newcommand{\MultOdds}{2.3\xspace}
\newcommand{\MultSD}{0.27\xspace}
\newcommand{\NDups}{135,185\xspace}
\newcommand{\NGames}{315\xspace}
\newcommand{\NMissPT}{25,682\xspace}
\newcommand{\NRaw}{21.75~million\xspace}
\newcommand{\NReviewers}{12.4~million\xspace}
\newcommand{\NReviews}{21.6~million\xspace}
\newcommand{\NSGapDiff}{$-$0.53\xspace}
\newcommand{\NSGapDiffSE}{0.16\xspace}
\newcommand{\NSaPost}{0.10\xspace}
\newcommand{\NSaPre}{0.70\xspace}
\newcommand{\NSbPost}{0.68\xspace}
\newcommand{\NSbPre}{0.74\xspace}
\newcommand{\NUpdates}{3,512\xspace}
\newcommand{\NatDiff}{$-$0.005\xspace}
\newcommand{\NatG}{2,342\xspace}
\newcommand{\NatHi}{0.018\xspace}
\newcommand{\NatLo}{$-$0.028\xspace}
\newcommand{\NatNPost}{145\xspace}
\newcommand{\NatNPre}{302\xspace}
\newcommand{\NewGapPost}{6.7\xspace}
\newcommand{\NewGapPre}{8.8\xspace}
\newcommand{\NewPost}{0.74\xspace}
\newcommand{\NewPre}{0.71\xspace}
\newcommand{\NewsDev}{29,279\xspace}
\newcommand{\NewsGames}{312\xspace}
\newcommand{\NewsItems}{44,421\xspace}
\newcommand{\ObsFix}{0.47\xspace}
\newcommand{\ObsFixHi}{0.84\xspace}
\newcommand{\ObsFixLo}{0.14\xspace}
\newcommand{\ObsNat}{0.83\xspace}
\newcommand{\ObsNatHi}{1.21\xspace}
\newcommand{\ObsNatLo}{0.55\xspace}
\newcommand{\OcB}{0.44\xspace}
\newcommand{\OcC}{0.45\xspace}
\newcommand{\OcCorrMc}{0.88\xspace}
\newcommand{\OcDiff}{0.008\xspace}
\newcommand{\OcDiffLoNinety}{$-$0.023\xspace}
\newcommand{\OcGames}{623\xspace}
\newcommand{\OcMatched}{5,719\xspace}
\newcommand{\OcTitles}{8,565\xspace}
\newcommand{\OrgPos}{85.9\xspace}
\newcommand{\OtATen}{0.23\xspace}
\newcommand{\OtAThirty}{0.32\xspace}
\newcommand{\OtAThree}{0.81\xspace}
\newcommand{\OtBTen}{0.10\xspace}
\newcommand{\OtBThirty}{0.15\xspace}
\newcommand{\OtBThree}{0.57\xspace}
\newcommand{\OtHiTen}{0.11\xspace}
\newcommand{\OtHiThirty}{0.12\xspace}
\newcommand{\OtLoTen}{$-$0.11\xspace}
\newcommand{\OtLoThirty}{$-$0.17\xspace}
\newcommand{\OtObsTen}{+0.01\xspace}
\newcommand{\OtObsThirty}{$-$0.03\xspace}
\newcommand{\OtObsThree}{$-$0.40\xspace}
\newcommand{\OtPredTen}{+0.02\xspace}
\newcommand{\OtPredThirty}{$-$0.01\xspace}
\newcommand{\OtPredThree}{$-$0.36\xspace}
\newcommand{\OtherExtra}{97.0\xspace}
\newcommand{\OtherOrg}{88.1\xspace}
\newcommand{\PFdpPost}{$-$2.2\xspace}
\newcommand{\PFdpPre}{$-$0.2\xspace}
\newcommand{\PPa}{$-$0.01\xspace}
\newcommand{\PPb}{$-$0.14\xspace}
\newcommand{\PPc}{0.02\xspace}
\newcommand{\PPd}{$-$0.40\xspace}
\newcommand{\PPe}{$-$0.45\xspace}
\newcommand{\PTaPost}{0.18\xspace}
\newcommand{\PTaPre}{0.73\xspace}
\newcommand{\PTbPost}{0.64\xspace}
\newcommand{\PTbPre}{0.81\xspace}
\newcommand{\PTlgPost}{0.26\xspace}
\newcommand{\PTlgPre}{0.39\xspace}
\newcommand{\PTmPost}{0.11\xspace}
\newcommand{\PTmPre}{0.50\xspace}
\newcommand{\PTshPost}{0.10\xspace}
\newcommand{\PTshPre}{0.49\xspace}
\newcommand{\PanDiff}{$-$0.38\xspace}
\newcommand{\PanF}{8\xspace}
\newcommand{\PanHi}{0.62\xspace}
\newcommand{\PanLo}{$-$4.49\xspace}
\newcommand{\PanN}{247\xspace}
\newcommand{\PhiNone}{0.34\xspace}
\newcommand{\PhiObs}{0.68\xspace}
\newcommand{\PhiPair}{0.59\xspace}
\newcommand{\PhiRel}{0.70\xspace}
\newcommand{\PhiRelHi}{1.09\xspace}
\newcommand{\PhiRelLo}{0.47\xspace}
\newcommand{\PlAbove}{+0.05\xspace}
\newcommand{\PlAboveB}{+0.04\xspace}
\newcommand{\PlAboveBSE}{0.05\xspace}
\newcommand{\PlAboveSE}{0.06\xspace}
\newcommand{\PlBelow}{+0.04\xspace}
\newcommand{\PlBelowSE}{0.10\xspace}
\newcommand{\PredFix}{0.52\xspace}
\newcommand{\PredNat}{1.08\xspace}
\newcommand{\PuCrPost}{1.59\xspace}
\newcommand{\PuCrPre}{0.79\xspace}
\newcommand{\PuDPPostAbs}{1.3\xspace}
\newcommand{\PuDPPreAbs}{6.7\xspace}
\newcommand{\PuDUPostAbs}{14.0\xspace}
\newcommand{\PuDUPreAbs}{16.4\xspace}
\newcommand{\PuDprimePost}{0.58\xspace}
\newcommand{\PuDprimePre}{0.96\xspace}
\newcommand{\PuFPost}{3\xspace}
\newcommand{\PuFPre}{10\xspace}
\newcommand{\PuHPost}{9\xspace}
\newcommand{\PuHPre}{38\xspace}
\newcommand{\PuNT}{40\xspace}
\newcommand{\PuSplits}{25\xspace}
\newcommand{\QMch}{+0.033\xspace}
\newcommand{\QMchSE}{0.012\xspace}
\newcommand{\RDBwHi}{7.1\xspace}
\newcommand{\RDBwLo}{6.4\xspace}
\newcommand{\RDDiffAbs}{6.6\xspace}
\newcommand{\RDDiffSE}{0.8\xspace}
\newcommand{\RDEq}{12.0\xspace}
\newcommand{\RDEqSE}{1.4\xspace}
\newcommand{\RDGM}{6.8\xspace}
\newcommand{\RDGMSE}{0.8\xspace}
\newcommand{\RDPostAbs}{7.4\xspace}
\newcommand{\RDPostSE}{0.7\xspace}
\newcommand{\RDPreAbs}{0.8\xspace}
\newcommand{\RDPreSE}{0.3\xspace}
\newcommand{\RatioDec}{5.54\xspace}
\newcommand{\RatioMax}{1.06\xspace}
\newcommand{\RatioMin}{0.91\xspace}
\newcommand{\RatioNov}{3.78\xspace}
\newcommand{\RatioOct}{1.39\xspace}
\newcommand{\RecGames}{37,610\xspace}
\newcommand{\RecRows}{41~million\xspace}
\newcommand{\RefAgePost}{886\xspace}
\newcommand{\RefAgePre}{623\xspace}
\newcommand{\RefBuyPost}{60\xspace}
\newcommand{\RefBuyPre}{70\xspace}
\newcommand{\RefGamesPost}{48\xspace}
\newcommand{\RefGamesPre}{78\xspace}
\newcommand{\RelN}{80\xspace}
\newcommand{\RelPost}{0.25\xspace}
\newcommand{\RelPre}{0.46\xspace}
\newcommand{\RocketPost}{24.1\xspace}
\newcommand{\SDaPost}{1.5\xspace}
\newcommand{\SDaPre}{6.2\xspace}
\newcommand{\SDbPost}{3.3\xspace}
\newcommand{\SDbPre}{6.2\xspace}
\newcommand{\SDsPost}{4.2\xspace}
\newcommand{\SDsPre}{6.6\xspace}
\newcommand{\SeedMax}{$-$0.36\xspace}
\newcommand{\SeedMin}{$-$0.40\xspace}
\newcommand{\SevDiff}{+4.7\xspace}
\newcommand{\SevDownPost}{11\xspace}
\newcommand{\SevDownPre}{42\xspace}
\newcommand{\SevHi}{7.2\xspace}
\newcommand{\SevLo}{2.7\xspace}
\newcommand{\SevPostAbs}{1.0\xspace}
\newcommand{\SevPreAbs}{5.7\xspace}
\newcommand{\SimBias}{0.05\xspace}
\newcommand{\StPostFa}{1\xspace}
\newcommand{\StPostHit}{5\xspace}
\newcommand{\StPreFa}{8\xspace}
\newcommand{\StPreHit}{34\xspace}
\newcommand{\StabLa}{0.78\xspace}
\newcommand{\StabPl}{0.83\xspace}
\newcommand{\SurgeAAbs}{10.8\xspace}
\newcommand{\SurgeASE}{2.1\xspace}
\newcommand{\SurgeAn}{87\xspace}
\newcommand{\SurgeBAbs}{3.5\xspace}
\newcommand{\SurgeBSE}{3.0\xspace}
\newcommand{\SurgeBn}{20\xspace}
\newcommand{\SurgePAbs}{4.1\xspace}
\newcommand{\SurgePSE}{1.1\xspace}
\newcommand{\SurgePn}{39\xspace}
\newcommand{\TFiveDid}{$-$2\xspace}
\newcommand{\TPa}{0.31\xspace}
\newcommand{\TPaSE}{0.016\xspace}
\newcommand{\TPb}{0.31\xspace}
\newcommand{\TPbSE}{0.014\xspace}
\newcommand{\TTenDid}{+6\xspace}
\newcommand{\TTwentyDid}{$-$3\xspace}
\newcommand{\UIVHi}{$-$0.002\xspace}
\newcommand{\UIVLo}{$-$0.614\xspace}
\newcommand{\UIVPost}{0.34\xspace}
\newcommand{\UIVPre}{0.67\xspace}
\newcommand{\UIso}{+0.105\xspace}
\newcommand{\UIsoHi}{0.251\xspace}
\newcommand{\UIsoLo}{$-$0.061\xspace}
\newcommand{\UMaPostAbs}{0.082\xspace}
\newcommand{\UMaPre}{$-$0.086\xspace}
\newcommand{\UMbPostAbs}{0.167\xspace}
\newcommand{\UMbPre}{$-$0.062\xspace}
\newcommand{\UMch}{+0.108\xspace}
\newcommand{\UMchSE}{0.045\xspace}
\newcommand{\UMddd}{+0.075\xspace}
\newcommand{\UMdddHi}{0.167\xspace}
\newcommand{\UMdddLo}{$-$0.026\xspace}
\newcommand{\UMgPost}{+0.085\xspace}
\newcommand{\UMgPostSE}{0.043\xspace}
\newcommand{\UMgPre}{$-$0.023\xspace}
\newcommand{\UMgPreSE}{0.018\xspace}
\newcommand{\UPct}{+0.7\xspace}
\newcommand{\UPctHi}{6.4\xspace}
\newcommand{\UPctLo}{$-$5.5\xspace}
\newcommand{\VCoPost}{0.80\xspace}
\newcommand{\VCoPre}{0.89\xspace}
\newcommand{\VDid}{+0.01\xspace}
\newcommand{\VDidHi}{0.07\xspace}
\newcommand{\VDidLo}{$-$0.04\xspace}
\newcommand{\VGames}{163\xspace}
\newcommand{\VPostA}{0.37\xspace}
\newcommand{\VPostB}{0.40\xspace}
\newcommand{\VPreA}{0.35\xspace}
\newcommand{\VPreB}{0.40\xspace}
\newcommand{\VSdPostA}{0.10\xspace}
\newcommand{\VSdPostB}{0.18\xspace}
\newcommand{\VSdPreA}{0.17\xspace}
\newcommand{\VSdPreB}{0.20\xspace}
\newcommand{\VSlPost}{0.47\xspace}
\newcommand{\VSlPre}{0.77\xspace}
\newcommand{\VarHi}{$-$0.63\xspace}
\newcommand{\VarLo}{$-$0.21\xspace}
\newcommand{\VarLoHi}{$-$0.49\xspace}
\newcommand{\VarLoLo}{$-$0.17\xspace}
\newcommand{\VarLoSigN}{13\xspace}
\newcommand{\VarN}{24\xspace}
\newcommand{\VarSigN}{22\xspace}
\newcommand{\VisLifeLo}{$-$0.28\xspace}
\newcommand{\VisLifePct}{$-$0.35\xspace}
\newcommand{\VisLoBase}{$-$0.28\xspace}
\newcommand{\VisLoCtrl}{$-$0.28\xspace}
\newcommand{\VisLoCtrlHi}{0.01\xspace}
\newcommand{\VisLoCtrlLo}{$-$0.61\xspace}
\newcommand{\VisN}{2,923\xspace}
\newcommand{\VisPctBase}{$-$0.34\xspace}
\newcommand{\VisPctCtrl}{$-$0.33\xspace}
\newcommand{\VisPctCtrlHi}{$-$0.14\xspace}
\newcommand{\VisPctCtrlLo}{$-$0.52\xspace}
\newcommand{\VisRhoAPost}{0.22\xspace}
\newcommand{\VisRhoAPre}{0.19\xspace}
\newcommand{\VisRhoBPost}{0.24\xspace}
\newcommand{\VisRhoBPre}{0.20\xspace}
\newcommand{\VisRhoDid}{$-$0.01\xspace}
\newcommand{\VisRhoHi}{0.20\xspace}
\newcommand{\VisRhoLo}{$-$0.19\xspace}
\newcommand{\VolLL}{4.96\xspace}
\newcommand{\VolPost}{4.25\xspace}
\newcommand{\VolPostHi}{4.77\xspace}
\newcommand{\VolPostLo}{3.70\xspace}
\newcommand{\VolPre}{0.97\xspace}
\newcommand{\WorstEta}{0.53\xspace}
\newcommand{\WorstEtaHi}{1.00\xspace}
\newcommand{\WorstEtaLo}{0.22\xspace}
\newcommand{\WorstExtra}{96.1\xspace}
\newcommand{\WorstExtraGap}{$-$0.9\xspace}
\newcommand{\WorstExtraGapSE}{0.3\xspace}
\newcommand{\WorstLoExtra}{0.28\xspace}
\newcommand{\WorstLoOrg}{0.53\xspace}
\newcommand{\WorstOrg}{81.4\xspace}
\newcommand{\WorstOrgGap}{$-$6.7\xspace}
\newcommand{\WorstOrgGapSE}{1.2\xspace}
\newcommand{\WorstPreA}{$-$14.0\xspace}
\newcommand{\WorstPreB}{$-$17.5\xspace}

%% file: sections/00_abstract.tex
\begin{abstract}
Platforms ask customers for online reviews to reduce self-selection, but theory conflicts on whether review solicitation makes online ratings more or less informative. Selection theory implies that asking helps, voice theory that it dilutes the news, and research on rating inflation that compression destroys information. We reconcile these views with signal detection theory, which separates what a rating knows (its sensitivity) from what it shows (its criterion). We test the framework on Steam, a leading PC video game store, whose client began in October 2019 to prompt players for a review after about three hours of play. Using a difference-in-discontinuities design at that inferred threshold in \NReviews Steam reviews of \NGames games, we find that review counts just above the threshold become \VolPost times those just below, and the jump in the positive share grows by \RDDiffAbs percentage points. Agreement with critics changes little near launch and erodes later. The rating's movement changes far more: pass-through from reviews written in the first two hours falls from \PTaPre to \PTaPost just above the threshold, against \PTbPre to \PTbPost just below, and a fixed rating tier responds to \DHaPost\% of downturns in those reviews above the threshold, down from \DHaPre\%, against \DHbPost\% and \DHbPre\% below it. What the stream knows changes less. At equal review counts its sensitivity falls more above the threshold, because added reviews respond about half as much and scores near the ceiling are noisier per review, but at natural counts added volume offsets this loss, leaving a change of \CorrNatDid relative to the stream below (95\% CI \CorrNatLo to \CorrNatHi). Failures that developers acknowledge, which run through complaints, still move the score after launch. Review prompts therefore mainly change how a rating must be read, and platforms that solicit reviews should recalibrate their rating labels.
\end{abstract}

\keywords{online reviews \and review solicitation \and rating inflation \and signal detection theory \and electronic word of mouth \and reputation systems \and platform design \and video games}

%% file: sections/01_intro.tex
\section{Introduction}
\label{sec:intro}

Digital platforms depend on online reviews and ratings. Buyers use them to choose \citep{chevalier2006}, sellers and developers watch them for early warnings, and platforms use them to decide what to promote: Valve tells Steam developers that a game whose score falls below 40\% is less likely to be featured \citep{valvedocs}. Yet most customers never write anything, so ratings are built from a selected few \citep{hu2009,schoenmueller2020}. The standard repair is to ask. On October 30, 2019, Valve released a new library for Steam in which the client asked players whether they would like to review a game they had recently played. Valve's stated goal was to ``make reviewing easier and include more voices in the system'' \citep{grayson2019}, and reviews across the store rose from about 17,000 a day to about 70,000 \citep{valve2020}.

Asking creates a design problem that managers have little basis to solve. A prompt changes who supplies a rating's data, while the rules that turn those data into a number and a label stay the same. A consumer uses a rating for two tasks: choosing among products, and noticing when a product has changed. The second task matters for software, games, apps and online services, which remain editable after sale \citep{kallinikos2013}, and Steam shows a 30-day score beside its lifetime score \citep{valvedocs}. A platform that adds a prompt therefore needs to know whether it has helped or harmed these tasks and, if it has harmed them, whether the remedy is to stop asking, to ask differently, or to change the display.

Existing theory gives conflicting answers. Theories of selection in online reviews imply that asking should help, because a prompt reaches customers who would otherwise stay silent and more reviews mean less noise \citep{hu2009,lihitt2008}. Theories of expectation disconfirmation \citep{oliver1980} and voice \citep{hirschman1970} imply that asking should hurt. People are more likely to review when their experience departs from what they expected \citep{ho2017}, and speaking up is a classic response to decline, so prompted reviews could dilute the news that voluntary reviews carry. Research on rating inflation implies harm by a different route: when scores crowd toward the top, they may say less \citep{filippas2022,aziz2023}. Each theory is well supported in its own domain, but none was built to explain what a prompt does to a rating's ability to signal change, and applied to that question they disagree.

The closest empirical work shows this tension without resolving it. \citet{askalidis2017} find that prompted ratings hold steady over time while self-motivated ratings trend downward, so the two streams have different temporal patterns. \citet{gao2025} find that solicitation crowds out organic reviews and shifts content from specific to abstract topics. Neither study tests how a prompt changes the ordering of products, a rating's ability to separate change from noise, and the movement of what is displayed, or explains when these outcomes diverge.

We argue that the conflict arises because all three theories treat a rating's informativeness as a single quantity. Signal detection theory separates two things that this treatment merges \citep{greenswets1966,swets1988}. \emph{Sensitivity} is how well a system can separate signal from noise, and the \emph{criterion} is how much evidence the system requires before it responds. A score can move less because the stream behind it carries less news, which is a loss of sensitivity, or because a bounded scale has compressed while a fixed rule still looks for the same movement, which raises the criterion. In plainer terms, the first concerns what a rating knows and the second what it shows. Viewed this way, each theory is right about a different force: selection about noise, voice about signal, and inflation about scale.

This argument leads to three research questions. First, does a usage-triggered review prompt preserve a rating's ordering of products? Second, when a rating moves less after a prompt, has it lost sensitivity, or has its display become more conservative? Third, which changes stay visible, and what can a display do about those that do not?

We answer these questions on Steam, where the prompt is triggered by usage and each public review records the playtime at which it was written. Crossing an inferred threshold at three hours of play with the launch date, in \NReviews reviews of \NGames games, we find that review counts just above the threshold become \VolPost times those just below, and that the jump in the positive share grows by \RDDiffAbs points. Near launch the rating's agreement with critics changes little, although it is lower by 2021--22.

The rating's movement changes far more. We measure it against short-playtime reviews, meaning those written in the first two hours of play. Pass-through, the share of a month's movement in these reviews that reaches another group of reviews, falls from \PTaPre to \PTaPost just above the threshold, compared with \PTbPre to \PTbPost just below it, and a rating tier with fixed percentage boundaries, like those of the store's labels, responds to \DHaPost\% of downturns just above the threshold, down from \DHaPre\% before launch, against \DHbPost\% and \DHbPre\% just below. What the stream knows changes less. At equal review counts, its correlation with short-playtime reviews falls by \CorrFixDidAbs more than the stream just below, because added reviews respond about half as much in log odds and a score near its ceiling is noisier per review. At natural counts that difference is \CorrNatDidAbs and is not distinguishable from zero, because more reviews offset this loss. Failures that developers acknowledge in their own announcements give evidence from outside the review data: they run through complaints, and the score over all reviews falls at them by \FaAllPostAbs points after launch, compared with \FaAllPreAbs points before, net of quiet periods in the same games.

Our contributions fall into two groups. On the theoretical side, we first reconcile three conflicting accounts of what solicitation does to ratings within one detection framework, showing that selection theory describes noise, voice theory describes signal, and inflation theory describes scale. Second, we extend signal detection theory from human and diagnostic judgments to a rating system, with a ratio model that gives closed-form expressions for responsiveness and sensitivity under stated conditions. Third, we identify a boundary condition, namely which flow carries a change: changes that run through complaints stay visible after a prompt, whereas changes in short-playtime reviews, which run partly through praise, fade. Fourth, we test a review-based signal against failures coded from developer announcements. Short-playtime reviews barely register those failures, so research that uses early reviews as a proxy for product change should test that link.

On the practical side, a platform that adds a usage-triggered prompt should expect its fixed labels to go quiet, responding less often to downturns and to noise alike, even where the review stream shows no detectable loss of information. Recalibration is a choice of operating point and not a free gain: on the reconstructed public score, no alternative rule we test beats the label at the label's own false-alarm rate, and rules of thumb learned before a prompt do not transfer. Reporting level and change from separate samples is a design hypothesis that the evidence supports testing. The conditions behind these results are not specific to games, since they require only that a product can change, that a prompt reaches people who stayed, and that fixed thresholds are used to read the score.

The rest of the paper follows this argument. Section~\ref{sec:literature} reviews the three theories and introduces signal detection, and Section~\ref{sec:theory} develops the research model and hypotheses. Sections~\ref{sec:data} and~\ref{sec:strategy} describe the setting, data and empirical strategy. Section~\ref{sec:results} reports the tests, Section~\ref{sec:discussion} draws out the implications for theory and design, and Section~\ref{sec:limitations} sets out limitations and future work.

%% file: sections/02_literature.tex
\section{Theoretical Background}
\label{sec:literature}

A rating system is an IT artifact \citep{orlikowski2001} with an input side and an output side. On the input side, the platform decides who is asked and when. On the output side, it decides how the resulting reviews become a number and a label. Research on online reviews has concentrated on the input. Our question concerns how a change at the input reaches a reader through an output rule that stays fixed. We first review three theories that bear on this question, and then introduce the framework we use to reconcile them.

\subsection{Selection: asking should repair the rating}

Online reviews, a central form of electronic word of mouth, matter for sales \citep{chevalier2006}, and through reputation systems they matter for trust on platforms \citep{tadelis2016}, yet they are built from a selected group. Reviews on most platforms pile up at the positive end of the scale, and platforms differ in polarity according to how selective customers are in reviewing \citep{hu2009,schoenmueller2020}. Self-selection makes the mean rating a biased estimate of quality \citep{hu2017}, silent customers can distort the public record \citep{dellarocas2008}, early buyers differ from later ones in ways that cost readers surplus when left uncorrected \citep{lihitt2008}, and learning from reviews is complicated because who buys and reviews depends on what earlier reviews said \citep{acemoglu2022}.

If selection is the problem, asking looks like the remedy, and a growing literature studies it. Financial incentives raise the number of reviews, whereas social norms raise their length \citep{burtch2018}. An unincentivized solicitation email reduces the extremity of ratings \citep{brandes2022}, solicitation draws in customers with moderate experiences and makes rating distributions more representative \citep{karaman2021}, and hiding reviews until both parties submit raises reviewing and reduces retaliation \citep{fradkin2021}. Applied to our question, this theory implies that added reviewers are more representative and otherwise similar to voluntary ones, and that a larger sample lowers noise. A rating should then rank products at least as well and catch change at least as soon.

\subsection{Disconfirmation and voice: asking should dilute the news}

A second body of work asks why people review, and it leads to a different expectation. A person is more likely to post when the gap between expected and experienced quality is larger \citep{ho2017}, which fits the long-standing view that satisfaction reflects disconfirmed expectations \citep{oliver1980}. Posting also responds to the ratings environment, and a core group of active reviewers can shape how opinion evolves \citep{moe2012}. In Hirschman's terms, customers respond to decline through exit, voice or loyalty \citep{hirschman1970}, and voice on digital platforms has been studied as a channel between consumers and firms \citep{kumar2018}.

This view matters because quality moves over time. Sequential and temporal dynamics in ratings are distinct processes \citep{godes2012}, aggregation can be improved by allowing for quality that changes \citep{dai2018}, and on eBay a seller's first negative feedback is a large event \citep{cabral2010}. Rating rules also feed back into how products change: Apple's App Store reset ratings at each update until 2017, which discouraged updating \citep{leyden2025}.

Applied to our question, this theory implies that voluntary reviews carry the news about change and that a prompted review, which needs no trigger, carries less of it. The evidence on prompted reviews is suggestive. Across four large retailers, email-prompted ratings run up to half a star above self-motivated web ratings and stay stable while web ratings trend down \citep{askalidis2016,askalidis2017}. Hotels that joined TripAdvisor's solicitation program saw a 34.3\% rise in review volume, a 0.151-star rise in ratings and a 16.9\% fall in review length, along with a 15.5\% fall in organic reviews and a shift from specific to abstract content \citep{gao2025}. Table~\ref{tab:positioning} summarizes the closest studies and their evidence on change over time.

\begin{table}[H]
\centering
\caption{The closest work on asking for reviews, and its evidence on change over time}
\label{tab:positioning}
\footnotesize
\begin{tabularx}{\textwidth}{>{\raggedright\arraybackslash}p{2.3cm} >{\raggedright\arraybackslash}p{2.6cm} >{\raggedright\arraybackslash}p{2.6cm} >{\raggedright\arraybackslash}p{2.9cm} >{\raggedright\arraybackslash}X}
\toprule
Study & Treatment & Design & Main outcomes & Evidence on change over time \\
\midrule
\citet{burtch2018} & Payments and social norms & Field and online experiments & Review volume and length & Not a focus \\
\citet{khern2018} & Monetary rewards on one platform & Difference in differences across platforms & Positivity, quality, participation & Not a focus \\
\citet{askalidis2017} & Email prompts at retailers & Prompted vs.\ self-motivated comparison & Rating level, selection & Prompted ratings stable, self-motivated ratings fall \\
\citet{karaman2021} & Solicitation by a firm & Private satisfaction ratings against public reviews & Extremity, representativeness & Not a focus \\
\citet{brandes2022} & Solicitation email & Field experiment & Extremity of ratings & Not a focus \\
\citet{jung2023} & Timing of reminders & Field experiments & Posting and review content & Timing of the request, not change in the product \\
\citet{gao2025} & Hotels opt into a solicitation program & Difference in differences across platforms & Volume, rating, length, organic reviews, topics & Crowding out and a shift toward abstract content \\
This study & Platform-wide, usage-triggered prompt & Difference in discontinuities across a playtime threshold & Ordering against critics; responsiveness; sensitivity; display & Whether movement and information change together, and which changes stay visible \\
\bottomrule
\end{tabularx}
\end{table}

\subsection{Rating inflation: compression should destroy information}

Where the first two theories concern who reviews, a third concerns what happens to a score once reviews crowd toward the top. Reputation inflation may make ratings less informative in the long run \citep{filippas2022}, and rating inflation, in which the average rises and the variance across listings falls, threatens informativeness even as it raises purchases \citep{aziz2023}. Applied to our question, this theory implies that a prompt, by lifting and compressing scores, makes a rating less informative. The argument merges a change of scale with a change of information. Compression on a bounded scale makes a given change in the balance of opinion smaller in percentage points and makes a fixed tier harder to cross, but neither fact shows that a reader who knew the new scale would learn less.

\subsection{Signal detection: separating what a system knows from what it shows}

These three theories point in different directions because each takes informativeness to be one quantity. Signal detection theory, which describes any system that must decide whether a signal is present in noise, separates two properties \citep{greenswets1966}. Sensitivity is how far apart signal and noise lie on the evidence the system receives, and it is summarized by the receiver operating characteristic, the trade-off between hits and false alarms as the system's threshold moves. The criterion is where on that evidence the system sets its threshold, and the point it reaches on the curve is its operating point. \citet{swets1988} shows that this framework applies to diagnostic systems of many kinds and that a measure based on the operating characteristic is uninfluenced by decision bias. Two systems can therefore have the same sensitivity and very different hit rates if one is more conservative than the other.

A rating system is such a detector. The signal is a change in how a product is experienced, the noise is ordinary variation in who reviews and what they say, and the output rule is a threshold, such as a fall of some size or the crossing of a label boundary. Treating the rating this way resolves the conflict among the three theories, because each concerns a different quantity. Selection theory is about noise: more reviews should raise sensitivity. Voice theory is about signal: unresponsive reviews should lower it. Inflation theory is about scale: compression leaves sensitivity alone in principle but moves a fixed rule to a more conservative point on its curve. Which force dominates each outcome is an empirical matter, and the next section develops a model that makes the outcomes separately testable.

%% file: sections/03_theory.tex
\section{Research Model and Hypotheses}
\label{sec:theory}

\begin{figure}[H]
\centering
\includegraphics[width=\textwidth]{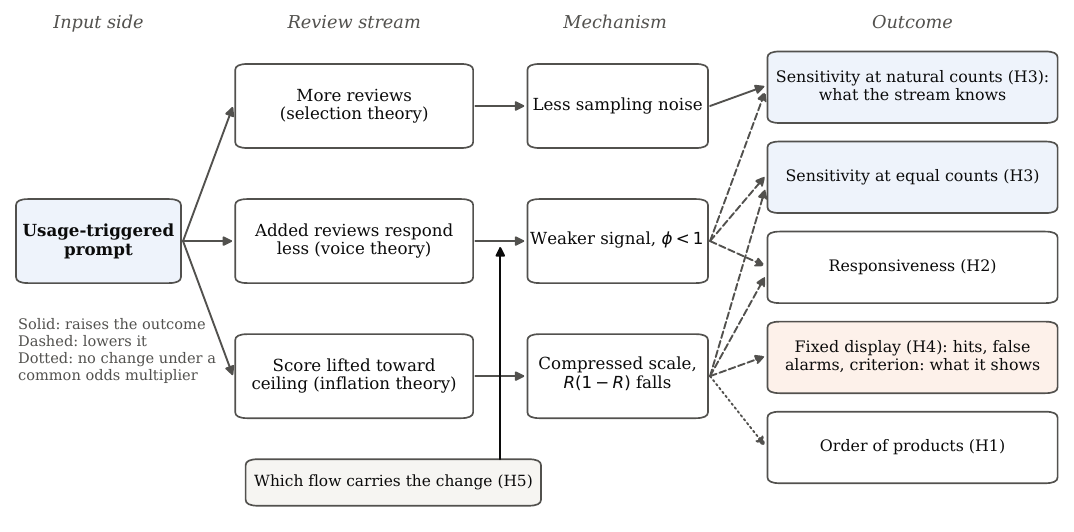}
\caption{Research model. A prompt on the input side of a rating system raises review volume, adds reviews that may respond less to change, and lifts the score toward its ceiling. The three mechanisms have different effects on what the stream knows and on what a fixed display shows.}
\label{fig:model}
\end{figure}

Figure~\ref{fig:model} summarizes our research model. A usage-triggered prompt changes three things about the review stream, and each works through its own mechanism: more reviews lower sampling noise, added reviews that respond less weaken the signal, and a higher score compresses the scale. Which flow carries a change moderates how much of the signal is diluted. We first define the outcomes, then formalize these mechanisms, and finally state our hypotheses.

\subsection{Outcomes}

Consider a score that reports the share of positive reviews of a product. For the task of choosing among products, the outcome is the \emph{order} of products. For the task of noticing change, three outcomes need to be kept apart. \emph{Responsiveness} is how far the score moves when the balance of opinion moves, on a stated scale. \emph{Sensitivity}, in the signal detection sense, is how well the stream separates periods of change from ordinary variation, whatever the scale; because a strictly increasing transformation of a fixed statistic leaves it unchanged, it measures what the stream knows. \emph{Display} is what a reader receives from the score and its label under fixed rules, which we measure by hits, false alarms and the criterion they imply, so it measures what the rating shows. These outcomes can diverge: a compressed score has lower responsiveness on the percent scale, and a fixed tier fires less often, even if a reader who knew the new scale could learn as much as before.

\subsection{A rating as a ratio of two flows}

To see how a prompt reaches these outcomes, write the score of product $g$ at time $t$ as a ratio of two flows, praise $L_{gt}$ and complaints $V_{gt}$:
\begin{equation}
R_{gt} = \frac{L_{gt}}{L_{gt} + V_{gt}}, \qquad x_{gt} \equiv \ln\frac{L_{gt}}{V_{gt}}, \qquad dR_{gt} = R_{gt}(1 - R_{gt})\, dx_{gt}.
\label{eq:ratio}
\end{equation}
The last expression is an identity. It says that a given change in the balance of praise and complaints moves the score by an amount proportional to $R(1-R)$, which shrinks as the score approaches its ceiling. A prompt adds $A_{gt}$ positive and $B_{gt}$ negative reviews, and we define the added shares $a = A/(L+A)$ and $b = B/(V+B)$. The first question is what this addition does to the order of products.

\begin{proposition}[Order]
If the prompt multiplies the odds of a positive review by a common factor $\kappa > 1$ across products, the order of products by score is unchanged, and differences in score shrink among products with scores above one half.
\end{proposition}

This proposition is a benchmark. Prompting does not imply a common factor, because products can differ in who reaches the trigger, who answers, and what the marginal respondent thinks, and different multipliers can reorder products. The question is therefore empirical, and we estimate how far the multipliers differ. The second question is what the addition does to the response of the score when a product changes.

\begin{proposition}[Responsiveness]
Let a change in the product move voluntary flows by $d\ln L = -\ell$ and $d\ln V = v$, with $\ell, v \geq 0$ and not both zero, and let added flows respond with elasticity $\eta \in [0,1]$ relative to voluntary ones. Let $w = \ell/(\ell + v)$ be the part of the change carried by praise. Relative to the response without the prompt, the percent response is scaled by a ceiling factor, $R'(1-R')/R(1-R)$, and a sensor factor,
\begin{equation}
\phi = w\,[1 - a(1-\eta)] + (1-w)\,[1 - b(1-\eta)].
\label{eq:phi}
\end{equation}
\end{proposition}

The sensor factor measures how much of the log-odds response survives dilution, and three cases matter. If added reviews respond exactly as voluntary ones do ($\eta = 1$), then $\phi = 1$ and only the ceiling factor remains, which is what selection theory implies. If added reviews respond less ($\eta < 1$) and enter a flow that carries the change, then $\phi < 1$, as voice theory implies, and $\phi$ lies between $1 - \max(a,b)(1-\eta)$ and $1 - \min(a,b)(1-\eta)$. A boundary case shows what behavioral evidence must establish: if a change runs almost entirely through complaints ($w \approx 0$) and few complaints are added ($b \approx 0$), then $\phi \approx 1$ even if added reviews do not respond at all. The sign restriction also matters, because when an event raises both flows, as an update that brings players back can, $w$ falls outside the unit interval and $\phi$ can exceed one. Log-odds evidence alone therefore cannot separate reviewers who respond less from a change that runs through a flow the prompt barely touched, so we estimate $a$, $b$ and $w$ directly. The third question is what these forces imply for the stream's ability to separate change from noise.

\begin{proposition}[Sensitivity]
Let the stream hold $n$ reviews before the prompt and $n'$ after, so that the sampling variance of the log odds is approximately $1/[nR(1-R)]$ before and $1/[n'R'(1-R')]$ after. Holding other sources of noise aside, the ratio of signal to noise in log odds changes by
\begin{equation}
\frac{D'}{D} = \phi \times \sqrt{\frac{n'\,R'(1-R')}{n\,R(1-R)}}.
\label{eq:detect}
\end{equation}
At equal counts ($n' = n$), $D'/D = \phi\sqrt{R'(1-R')/R(1-R)} < 1$ whenever the prompt raises a score above one half and $\phi \le 1$. At natural counts ($n' > n$), added volume works in the opposite direction, and the net effect has no fixed sign.
\end{proposition}

Proofs are in Appendix~\ref{app:model}. Taken together, the propositions separate three channels. The ceiling lowers percent responsiveness and raises the log-odds noise per review, dilution lowers the signal, and volume lowers the noise. A fixed tier on the percent scale sees the first two and gets little help from the third, because a tier responds to the size of the movement. Binomial noise is not the only noise in a stream, so Equation~\eqref{eq:detect} gives an upper limit on what added volume can buy.

\subsection{Why added reviews might respond less}

The model takes the response of added reviews, $\eta$, as given, and Hirschman's framework offers a useful lens for what to expect. A voluntary reviewer is often using voice, and because people post when their experience departs from their expectation \citep{ho2017}, a change in the product should raise voluntary posting and shift its content. A player who is hit hard by a change may exit and so never be asked. A prompted review comes from someone who reached the trigger and was handed a low-cost question. We expect that review to reflect stable fit with the game more than recent change, which implies $\eta < 1$. We do not observe exit, loyalty or fit directly, and a positive review is not itself evidence of loyalty, so the prediction we test is the one that can be measured: added reviews respond less than voluntary reviews near the trigger.

A second route leads to the same prediction. A prompted reviewer answers inside the client and need not pass through the store page, where the score is displayed. A voluntary reviewer who has just read that score may lean toward it, so part of the co-movement among voluntary reviews can reflect the displayed score and not the reviewer's own news. Losing that channel would also lower $\eta$, with a different meaning: the rating would lose an echo and not a signal. We test this route directly in Section~\ref{sec:added}.

\subsection{Hypotheses}

Table~\ref{tab:theories} sets out what each theory implies when applied to a usage-triggered prompt, and what the detection view adds. These are our readings of the theories for this setting, since none of them addresses it directly. We begin with the outcome on which the theories differ least. The common-multiplier benchmark, together with the persistence of fit, suggests that the order of products should change little at first.

\begin{table}[H]
\centering
\caption{What each theory implies for a usage-triggered prompt}
\label{tab:theories}
\footnotesize
\begin{tabularx}{\textwidth}{>{\raggedright\arraybackslash}p{2.1cm} >{\raggedright\arraybackslash}X >{\raggedright\arraybackslash}X >{\raggedright\arraybackslash}X >{\raggedright\arraybackslash}X >{\raggedright\arraybackslash}X}
\toprule
Theory & Order of products & Responsiveness & Sensitivity per review & Sensitivity of the stream & Fixed display \\
\midrule
Selection & Holds or improves & Falls only through the ceiling & Unchanged & Rises & Holds \\
Disconfirmation and voice & Holds & Falls beyond the ceiling & Falls & Falls & Loses hits \\
Rating inflation & Worsens & Falls & Falls & Falls & Loses hits \\
Detection view & Holds near launch & Falls through ceiling and dilution & Falls & Ambiguous: volume offsets dilution & Loses hits and false alarms \\
\bottomrule
\end{tabularx}
\end{table}

\textbf{H1 (Order).} A prompt raises the level of ratings and compresses their spread, and near its launch the rating's agreement with independent measures of quality stays within a small margin.

The theories split on responsiveness in a way the log-odds scale can test, because the ceiling acts only on the percent scale. Selection theory expects the loss to disappear on that scale, whereas voice theory expects it to remain.

\textbf{H2a (Selection).} After a prompt, percent pass-through falls just above the usage trigger relative to just below it, and log-odds pass-through does not.

\textbf{H2b (Voice).} After a prompt, pass-through falls just above the usage trigger relative to just below it on the log-odds scale as well as the percent scale.

They split on sensitivity as well. If added reviews lower the signal, voice and inflation accounts imply a loss of information in the stream, whereas the detection view implies a loss per review that volume can offset.

\textbf{H3a (Information loss).} Relative to the stream just below the usage trigger, the stream just above it has lower sensitivity at natural review counts.

\textbf{H3b (Offsetting forces).} Relative to the stream just below the usage trigger, the stream just above it has lower sensitivity at equal review counts, and the difference is smaller or absent at natural counts.

The detection view also makes a prediction about the display. A compressed score should cross a fixed boundary less often, so a fixed rule should behave more conservatively even where the stream retains its information.

\textbf{H4 (Display).} Fixed rating tiers lose hits above the usage trigger relative to below it, and they become more conservative. Thresholds set for the new scale recover part of the loss in hits.

Finally, the boundary case of Proposition~2 implies that the channel of a change matters. A change that raises complaints should remain visible even when praise has been inflated.

\textbf{H5 (Channel).} Changes that run through complaints remain visible in the score after a prompt.

%% file: sections/04_setting.tex
\section{Setting and Data}
\label{sec:data}

We test these hypotheses on Steam, which offers a usage-triggered prompt, a precise record of when each review was written, and a long panel of products that change after sale.

\subsection{Steam and the prompt}

Steam sells PC games and hosts a review system in which a player who owns a game recommends it or not and may add text. Each review displays the player's playtime when it was written. The store page shows two scores, one over the past 30 days and one over the product's lifetime, and only reviews from purchases on Steam count toward them \citep{valvedocs}. Each score appears with a label. Valve confirms that 40\% separates ``Mixed'' from ``Mostly Negative'' \citep{valvedocs} but does not publish the other rules, so we test them against the labels the store displays. The labels change at 95, 80, 70, 40 and 20\% positive, with count conditions at 50 and 500 reviews: ``Overwhelmingly Positive'', for example, needs 95\% and at least 500 reviews, and a score of 95\% on fewer reviews shows as ``Very Positive''. These rules reproduce the displayed label for all \LabelN games with at least 10 reviews that we queried through Steam's public store interface on \LabelDate, a set that includes every game in our main sample. Games are updated frequently after release through patches, new content and changes to their business model, which is why the 30-day score matters to a buyer.

The prompt arrived with the new library on October 30, 2019. Valve described it at the time: ``When a player visits the app page in the new library for a recently played game they have not yet reviewed, the Steam client will ask them whether they would like to leave a review'' \citep{grayson2019}. Valve later reported that the new ways to prompt users to review or re-review a game raised daily reviews by over 300\% \citep{valve2020}. Its documentation states that users are prompted ``after playing a game for a period of time'' and that a user who left a negative review and kept playing may be prompted to update it \citep{valvedocs}. Because the question is answered inside the client, the reviewer need not open the store page, where the scores appear. Valve does not publish the trigger, so we infer it from outcomes. As we show below, the first threshold sits at 180 minutes of play, with further thresholds at 10 and 30 hours and at higher milestones. No review carries a flag, so the threshold marks inferred eligibility and not observed receipt of a prompt.

\subsection{Data}

Our main data are a public dataset of Steam reviews posted to Kaggle in January 2021 by an independent collector \citep{najzeko2021}. It holds \NRaw rows for \NGames popular games selected by the collector, posted from November 2010 to January 23, 2021. Each record gives the game, date, recommendation, text, playtime at the time of the review, later playtime, purchase channel, helpful votes, and counts of the reviewer's games and reviews. We drop \NMissPT rows with no playtime at review and \NDups duplicate rows, nearly all from one game that appears twice, which leaves \NReviews reviews by \NReviewers accounts. We compute text features and then discard the text. The pre-period runs from January 2017 to September 2019 and the post-period from December 2019 to January 2021, with January 2021 a partial month, and we exclude October and November 2019 as the rollout months.

Because the dataset is a snapshot, a recommendation that was edited before January 2021 appears in its edited form under its original date. In the study window, \EditPost\% of post-launch reviews and \EditPre\% of pre-launch reviews carry a later update time, and we report results for never-edited reviews below. Table~\ref{tab:summary} describes the sample.

\input{generated/tab_summary}

The headline count of reviews is not the unit of most of our tests. Our main comparison uses game-months, meaning one game in one calendar month. The \emph{paired sample} holds the game-months in which both bands around the threshold meet a minimum review count: \FlowPairPreGM game-months across \FlowPairPreG games before launch and \FlowPairPostGM across \FlowPairPostG games after it. Table~\ref{tab:flowchart} traces how the sample narrows at each filter. Because minimum counts can select different game-months once a prompt raises volume, we also report estimates on the \FlowBothG games observed in both periods.

\input{generated/tab_flowchart}

We add three further sources. For tests that compare games, a second, store-wide public dataset provides \RecRows dated recommendations across \RecGames games, running to December 2022 \citep{kozyriev2024}. It records total hours and not playtime at review, so we use it only for between-game comparisons, in which every period is sampled to the same number of reviews per game. As independent benchmarks of quality, we use Metacritic scores from a public snapshot of the Steam store \citep{sidorowicz2026} and top-critic averages from a public OpenCritic file \citep{rivers2025}, matched to Steam on normalized titles. Of \OcTitles OpenCritic titles with a score, \OcMatched match a Steam title, and the two critic measures have a rank correlation of \OcCorrMc. Critic scores are mostly fixed near release, so they benchmark a persistent ordering of games and do not track current quality after substantial updates. Finally, we pulled the news record that Steam holds for each game, covering \NewsItems items across \NewsGames games, of which \NewsDev are developer announcements. We use only developer announcements, because timestamps on third-party press items in this feed can be unreliable, and from them we build two event sets, described in Section~\ref{sec:strategy}: \NUpdates dated updates from a keyword rule and \FailEpisodes failure episodes coded from the text of the announcements.

One feature of the record requires care. Since March 2019, Valve has removed from the score periods it judges to be off-topic review bombs, meaning bursts of reviews aimed at lowering a score, on a topic that Valve considers unrelated to whether future buyers will be happy with the game \citep{valve2019,valvedocs}. Such periods also appear to be absent from the dataset: Rocket League, for example, has no reviews for May 1--9, 2019, the days after its studio's sale to Epic Games. We find \HoleDays such days across \HoleGames large games, all after the filter began. Removing a calendar interval from every band helps comparisons within a game-month, but it need not cancel if bands differ in timing and composition. For comparisons across time, we therefore also report a pre-period of March to September 2019, which falls under the same filter, and we report estimates without the affected games.

\subsection{Review groups}

We refer to reviews by the playtime at which they were written. \emph{Short-playtime reviews} are those written below two hours, which sit below the inferred threshold for this library prompt, and the \emph{short-playtime score} is their positive share. The \emph{adjacent bands} are 2--3 hours and 3--4 hours, on either side of the threshold, so they are close in playtime but not identical. Reviews above three hours after launch are in the \emph{prompted range}, although not all of them are prompted. \emph{Added reviews} are the additional reviews near the threshold after launch, relative to the flow just below it, and we infer their properties from the jump under assumptions we state and vary below. The next section uses these groups to test each hypothesis in turn.

%% file: generated/tab_summary.tex
\begin{table}[H]
\centering
\caption{The review sample before and after the prompt}
\label{tab:summary}
\small
\begin{tabular}{lrr}
\toprule
Measure & January 2017--September 2019 & December 2019--January 2021 \\
\midrule
Reviews & 7,621,883 & 9,220,948 \\
Games & 292 & 315 \\
Calendar months & 33 & 14 \\
Positive reviews (\%) & 81.0 & 91.4 \\
Median playtime at review (hours) & 36.7 & 26.4 \\
Reviews by players under 3 hours (\%) & 10.5 & 5.8 \\
Median review length (characters) & 53 & 24 \\
Reviews of 200+ characters (\%) & 22.7 & 11.5 \\
Reviews with a helpful vote (\%) & 39.0 & 16.7 \\
Reviewer has one review on record (\%) & 18.8 & 27.5 \\
Bought on Steam (\%) & 76.1 & 80.4 \\
Written in English (\%) & 41.0 & 43.6 \\
\midrule
\multicolumn{3}{l}{\textit{Short-playtime reviews (under 2 hours)}} \\
\quad Reviews & 590,963 & 398,390 \\
\quad Positive (\%) & 65.3 & 78.1 \\
\quad Written in English (\%) & 38.1 & 38.2 \\
\quad Bought on Steam (\%) & 69.8 & 59.7 \\
\quad Reviewer has one review on record (\%) & 16.1 & 16.6 \\
\quad Median games owned by reviewer & 78 & 48 \\
\quad Median days since the game's first review & 623 & 886 \\
\bottomrule
\end{tabular}
\par\vspace{2pt}{\footnotesize\raggedright October and November 2019 are excluded as the rollout months. Review length and helpful votes are measured at the time the data were collected, in January 2021. January 2021 ends on the 23rd.\par}
\end{table}

%% file: generated/tab_flowchart.tex
\begin{table}[H]
\centering
\caption{From reviews to the units of the main tests}
\label{tab:flowchart}
\small
\begin{tabular}{lrrrr}
\toprule
Step & \multicolumn{4}{r}{Reviews} \\
\midrule
Raw rows in the dataset & \multicolumn{4}{r}{21,747,371} \\
Reviews after removing rows with no playtime and duplicates & \multicolumn{4}{r}{21,586,504} \\
\quad before the pre-period & \multicolumn{4}{r}{3,564,330} \\
\quad rollout months & \multicolumn{4}{r}{1,179,343} \\
\midrule
 & \multicolumn{2}{c}{Before launch} & \multicolumn{2}{c}{After launch} \\
 & Game-months & Games & Game-months & Games \\
\midrule
Game-months with any review & 7,354 & 292 & 4,279 & 315 \\
Short-playtime reviews 30+ & 3,261 & 270 & 1,798 & 206 \\
\quad and 2--3 hour band 20+ & 2,174 & 241 & 1,242 & 167 \\
\quad and 3--4 hour band 20+ & 2,016 & 230 & 1,747 & 198 \\
\quad and both bands 20+ (paired sample) & 1,878 & 222 & 1,241 & 167 \\
Paired sample, games in both periods & 1,667 & 143 & 1,090 & 143 \\
\midrule
Paired sample, median reviews: short-playtime / 2--3 h / 3--4 h & \multicolumn{2}{r}{142 / 52 / 47} & \multicolumn{2}{r}{123 / 46 / 176} \\
\bottomrule
\end{tabular}
\end{table}

%% file: sections/05_strategy.tex
\section{Empirical Strategy}
\label{sec:strategy}

Our strategy follows the hypotheses in order. We first establish what the prompt did to the review stream, and we then define the estimators for order, responsiveness, sensitivity, display and channel. We close with the out-of-sample check of the model.

\subsection{The review stream at the threshold}

Playtime at review is observed only for people who choose to review, and the prompt changes that choice. The threshold therefore compares review populations, not otherwise identical individuals with and without treatment. Our estimand is the local change in the review stream around an inferred eligibility boundary: its count, composition and positive share. The jump in counts is an outcome here and the evidence of solicitation, so the usual test for no jump in density does not apply.

We estimate a difference in discontinuities, which compares the jump at the threshold after launch with the jump before launch \citep{grembi2016}. For review $i$ of game $g$ at playtime $p_i$, separately before and after launch, we estimate a local linear regression around the 180-minute cutoff \citep{lee2010}:
\begin{equation}
y_{ig} = \alpha_g + \tau\, \mathbf{1}[p_i \geq 180] + \beta_1 (p_i - 180) + \beta_2 (p_i - 180)\mathbf{1}[p_i \geq 180] + \varepsilon_{ig},
\label{eq:rd}
\end{equation}
with a 30-minute bandwidth. The two periods are estimated jointly, so the change in $\tau$ has a clustered standard error with its cross-period covariance. As sensitivity checks, we absorb game-by-month strata, give each game-month equal weight, and vary the bandwidth.

Interpreting the change in $\tau$ as the effect of the prompt requires three assumptions: that counts and composition would have been smooth through the boundary without the prompt, up to a stable discontinuity; that nothing else specific to this threshold changed at the launch; and that the launch changed the stream near the threshold through the prompt as delivered, a question answered inside the client. The treatment therefore bundles being asked with answering away from the store page, and we test the second part in Section~\ref{sec:added}. Because we located the threshold from the data, we also check it out of sample. We search for the largest jump in a random half of games and test it in the other half, and we search in one block of time and test it in another.

\subsection{Order (H1)}

We measure a rating's ordering of products as the Spearman rank correlation across games between the positive share and a critic benchmark. Our main estimand holds sampled review counts constant: in each period we draw the same number of reviews per game, compute the correlation within that draw, repeat 40 times, and average. The change across the launch and its interval come from a bootstrap of games that recomputes this same statistic in every period. Equal counts equalize the number of reviews but not all sampling noise, because the variance of a share also depends on its level. We also report a natural-volume comparison that uses every review in the store-wide dataset in each period.

Since a failure to reject zero does not establish preservation, we use a noninferiority margin, the largest fall we treat as negligible: \Margin in Spearman's rho. For the pre-launch correlation, a fall of that size means the score and critics agree on \AgreeCost fewer of every 100 pairs of games, and in the same games it is about what the correlation loses when reviews per game fall from 100 to 25 (\LostHundred to \LostTwentyFive). We call the change noninferior when the lower bound of the 90\% interval lies above $-$\Margin. We also compare adjacent bands, measure rank stability across the launch against an earlier boundary, and test how far the odds multiplier in Proposition~1 differs across games.

\subsection{Responsiveness (H2)}

Turning to the second task, we measure responsiveness by how much of a month's movement in short-playtime reviews reaches each band. For band $v$ in game $g$ and month $m$, we estimate
\begin{equation}
s^{v}_{gm} = \mu_g + \lambda_m + \pi^{v}\, s^{\text{short}}_{gm} + \epsilon_{gm},
\label{eq:pass}
\end{equation}
where $s$ denotes the positive share and $s^{\text{short}}$ is the short-playtime score. We call $\pi^{v}$ the \emph{pass-through}. Game effects absorb a game's persistent quality, so the coefficient captures co-movement within a game, and month effects absorb shocks common to all games in a month. The coefficient identifies co-movement with short-playtime reviews, which need not reflect product change, so we test that link separately with independent events below.

Because $s^{\text{short}}_{gm}$ is measured with sampling noise, ordinary least squares would attenuate $\pi$, and by different amounts in periods with different counts. We therefore treat random halves of the short-playtime reviews in a game-month as repeated measures of the same latent share and use one half to instrument the other \citep{gillen2019}. Given the latent share, independent reviews make the halves' sampling errors independent, so the ratio of covariances recovers $\pi$ (Appendix~\ref{app:iv}). This step removes reviewer-level sampling noise; it does not make the latent share exogenous. Point estimates average over 40 random splits.

We estimate the equation on the percent and log-odds scales, because H2a and H2b differ on the second. For inference, we resample whole game histories, recompute the full estimator in every period, and take the contrast within each replicate, so that cross-period dependence and split variation are both reflected. We report first-stage $F$ statistics after fixed effects, sample counts, and sensitivity to the split seed.

The sharpest comparison subtracts the band just below the threshold from the band just above it, in the same game and month:
\begin{equation}
s^{3\text{--}4h}_{gm} - s^{2\text{--}3h}_{gm} = \mu_g + \lambda_m + \gamma\, s^{\text{short}}_{gm} + \epsilon_{gm}.
\label{eq:gap}
\end{equation}
The coefficient $\gamma$ is the \emph{pass-through gap}. The identifying assumption is that, without the prompt, players on either side of three hours would have moved alike with the short-playtime score. We check three implications of this assumption: $\gamma$ should be near zero before launch, should change at the launch date and not at earlier dates, and should change at the 180-minute cutoff and not at others.

\subsection{Sensitivity (H3)}

Sensitivity must be tested on a scale-free measure and with the same comparison months for each stream, so we use the paired sample. Our first measure is the correlation of a band's log odds with the short-playtime score, within game and month, over all these months. Our second is rank separation, the area under the receiver operating characteristic (AUC). It is the chance that a randomly chosen \emph{turn month}, in which the short-playtime score falls ten or more points below its trailing median, shows a larger fall in the band than a randomly chosen \emph{calm month}, in which that score stays within five points of its median. Trailing medians use the previous six calendar months within a regime, where regimes are the periods separated by Valve's March 2019 filter and by the launch, and they require at least three eligible months. Because AUC depends only on ranks, it is invariant to an increasing transformation of a fixed statistic. A change from a baseline is not such a statistic, so we report it on the log-odds scale and check the percent scale.

We compute each measure at natural counts and at equal counts, drawing the same number of reviews from each band in every month. Natural counts measure actual platform performance, and equal counts isolate composition from volume. The contrast of interest, which we call the \emph{threshold-specific change}, is the change above the threshold minus the change below it, because any shift in how often the short-playtime score carries real news affects both bands.

\subsection{Display (H4)}

We keep two analyses of the display apart. The first uses a \emph{diagnostic score}, built from adjacent bands or from reviews above three hours bought on Steam, on which we apply the percentage boundaries of the store's labels and call the resulting categories \emph{tiers}. The adjacent bands are read against the short-playtime score, and the score above three hours against reviews under three hours. Neither shares reviews with its reference, which makes the score suitable for inference, but it is a component of the review stream and not what buyers saw.

The second uses a \emph{reconstructed public score}: rolling 30-day windows ending on the last day of each month, Steam purchases only, all playtimes, with the store's labels and their count conditions. It approximates the historical score, because edits and deletions before the snapshot cannot be undone. Since this score contains short-playtime reviews, we prevent shared reviews from creating a mechanical link. We compute the short-playtime score from a random half of those reviews, remove that half from the public score, and count the other half twice. Results average over random halves, contrasts come from a game bootstrap that redraws the half, and results without this correction are nearly the same.

A \emph{hit} is a detection in a turn month, and a \emph{false alarm} is a detection in a calm month, where calm is defined by short-playtime reviews alone. From the hit rate $H$ and the false-alarm rate $F$ we compute the standard detection measures, sensitivity $d' = z(H) - z(F)$ \citep{greenswets1966} and criterion $c = -\tfrac{1}{2}[z(H) + z(F)]$, where $z$ is the normal quantile and a higher $c$ means a more conservative rule \citep{hautus2021}, with a correction for extreme rates \citep{hautus1995}. To compare displays we use matched false-alarm rates: thresholds are set on calm months in other games and evaluated on held-out games, within a period.

\subsection{Independent events (H5)}

To test the channel, and to validate the short-playtime score itself, we use two event sets that owe nothing to review scores. We use them to validate detection, not to estimate the causal effect of an event.

The first set consists of documented failures, meaning failures that developers acknowledge in their own announcements. A fixed keyword rule selects \FailCand developer announcements that may acknowledge a problem, such as outages, rollbacks, apologies, compensation or lost progress. Each was coded in shuffled order, with no review data and no date or game field, under a written codebook as a major failure, a limited failure, or neither, and as technical or a decision, with direction and severity taken only from the text (Appendix~\ref{app:coding}). Announcements about one failure are merged into an episode if they fall within 14 days in the same game, which gives \FailEpisodes episodes (\FailMajor major) across \FailGames games. We fix the window on pre-launch data alone. Reviews from 42 to 7 days before the announcement form the baseline, and reviews from 7 days before to 14 days after form the event window, because the pre-launch fall begins in the week before the announcement. We compare each band's change with its change in \emph{quiet windows}, which are dates in the same games with no failure within 42 days, and each window has equal weight.

The second set consists of dated updates. A news item counts as a developer update if it passes fixed title rules (Appendix~\ref{app:rules}), and events in a game are at least 14 days apart. For each update, we compare the 28 days before with the 14 days after, for reviews at 1--3 hours and at 3--5 hours of play, with the whole window inside one period. A pooled Mantel--Haenszel log odds ratio handles sparse cells without smoothing \citep{mantel1959}. An update is an intervention and not a known deterioration, since updates differ in severity and direction and developers may issue them in response to existing problems. Because spacing events 14 days apart leaves other updates in some baselines, we report isolated updates as a sensitivity check. Quiet windows provide a placebo, and all contrasts come from one game-level bootstrap.

\subsection{Out-of-sample check of the model}

We developed the model with the three-hour results in view, so its fit at three hours is a description and not a test. The prompts at 10 and 30 hours played no part in building it. Far fewer reviews are added there relative to the band below, so the model implies much smaller threshold-specific changes. We hold the part of a change carried by praise and the response of added reviews at their three-hour values, compute the model's implied changes from flows, scores and counts at each prompt, and record them before computing the outcomes at those prompts. The one exception is the percent gap at 10 hours, which had already appeared among the placebo cutoffs.

%% file: sections/06_results.tex
\section{Results}
\label{sec:results}

We first establish that the prompt changed the review stream, and we then take the hypotheses in turn. Section~\ref{sec:summary} collects the verdicts.

\subsection{What changed in the review stream}

\begin{figure}[H]
\centering
\includegraphics[width=\textwidth]{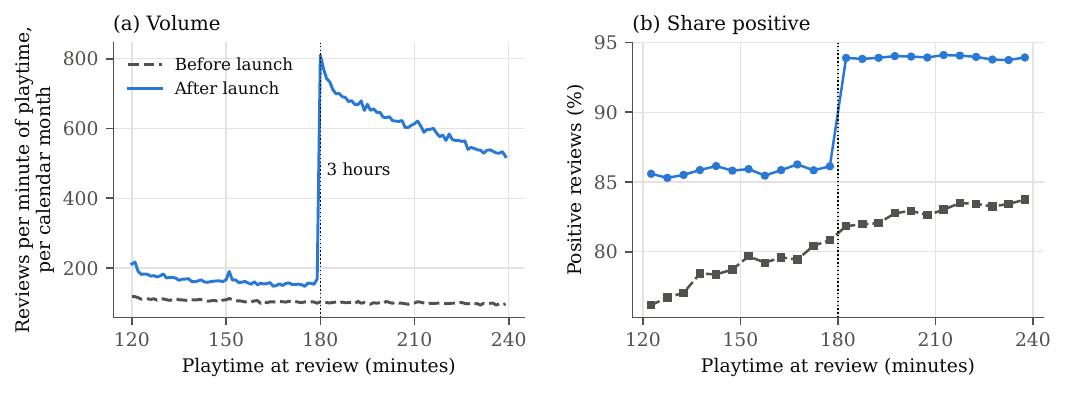}
\caption{Reviews and their positive share around three hours of play, before and after the prompt launched. Volume is scaled per calendar month to make the two periods comparable.}
\label{fig:threshold}
\end{figure}

\begin{figure}[H]
\centering
\includegraphics[width=\textwidth]{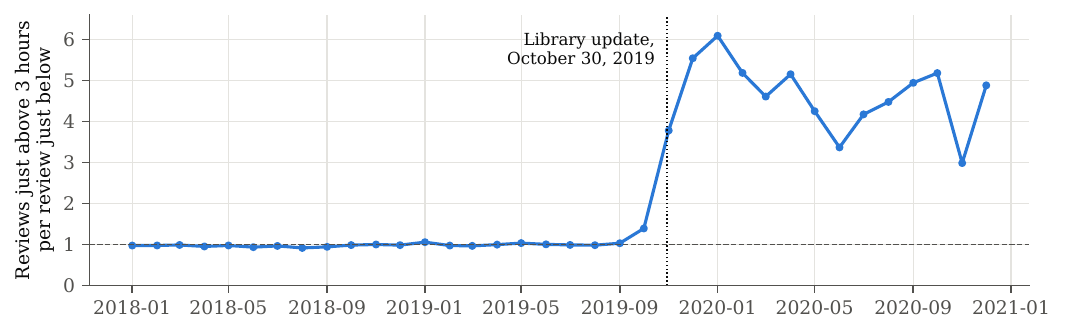}
\caption{Reviews written just above three hours of play per review just below, by month ($\pm$20 minutes).}
\label{fig:monthly}
\end{figure}

The prompt left a clear mark on the review stream, both at the threshold and at the launch date. After launch, reviews at 180 minutes of play outnumber those at 179 minutes by \MinAfter to \MinBefore, whereas before launch the same two minutes have \MinAfterPre and \MinBeforePre (Figure~\ref{fig:threshold}). Over a 30-minute window on each side, reviews above the threshold outnumber those below by \VolPost to one after launch (95\% CI \VolPostLo--\VolPostHi), against \VolPre before it, and local linear fits extrapolated to the cutoff give a ratio of \VolLL. The timing is equally sharp: the ratio stays between \RatioMin and \RatioMax in every month from January 2018 to September 2019, and it then moves to \RatioOct in October, \RatioNov in November and \RatioDec in December 2019 (Figure~\ref{fig:monthly}).

Because we located the threshold from the data, we also test it out of sample. In \HoldSplits random splits of games, a search for the largest count jump between 1 and 12 hours in one half picks 180 minutes in \HoldShare\% of splits. In the other half, the ratio at the chosen cutoff averages \HoldVal, against \HoldOther at cutoffs more than 30 minutes away. A search on December 2019 to June 2020 picks \HoldEarly minutes, and the ratio at that cutoff from July 2020 on is \HoldLate, against a median of \HoldLateMed across all cutoffs.

The added volume comes with a more positive stream. The jump in the positive share is \RDPostAbs points after launch (SE \RDPostSE) and \RDPreAbs before (SE \RDPreSE), a change of \RDDiffAbs points (SE \RDDiffSE; Table~\ref{tab:prompt}). We treat this change, and not the post-launch discontinuity itself, as our estimate of the effect at the threshold. It ranges from \RDBwLo to \RDBwHi across bandwidths of 10 to 60 minutes and is \RDGM (SE \RDGMSE) with game-by-month strata. With equal weight per game-month, which raises the influence of small games, it is \RDEq (SE \RDEqSE). The reviews just above the threshold also differ in other ways: they are shorter and attract fewer helpful votes, and their authors have less history on the platform and are more likely to keep playing.

\input{generated/tab_prompt}

These jumps allow us to characterize the added reviews. If voluntary reviewing is smooth across the threshold, the added reviews are \ExtraPos\% positive, against \OrgPos\% just below. Because this figure is a residual, we vary what it assumes (Table~\ref{tab:mixture} in the Appendix). It is \MixTen to \MixSixty across bandwidths of 10 to 60 minutes, \MixLL with local linear flows, \MixPre after carrying the pre-launch discontinuities into the counterfactual, and \MixGM with equal weight per game-month. Letting the voluntary flow above the threshold differ from the flow below by up to 50\%, and letting voluntary reviews above be up to 6 points more positive, gives a range of \GridMin to \GridMax. The prompt may also move reviews from later playtime to earlier, in which case these reviews are additional at this point in playtime without being additional overall. Counts at 3--10 and 10--30 hours rise in the same games (Table~\ref{tab:flow}), which leaves no visible displacement at later playtime.

The threshold at three hours is only the first. Counts jump \CutTen-fold at 10 hours and \CutThirty-fold at 30 hours after launch, with smaller spikes at higher milestones and a \CutThou-fold jump at 1,000 hours, and none of these jumps exists before launch (Table~\ref{tab:cutoffs} in the Appendix).

Praise also rose far more than complaints. In the \FlowGames games observed in both periods, positive reviews per game per day rose \FlowPosX-fold and negative reviews \FlowNegX-fold (Table~\ref{tab:flow}). Under three hours, positive reviews changed by a factor of \FlowUPos and negative reviews by \FlowUNeg, whereas in the 3--10 hour band positive reviews rose \FlowTPos-fold and negative reviews \FlowTNeg-fold. Reviews of 200 or more characters also rose in the prompted range, \FlowLongT-fold at 3--10 hours, while under three hours they changed by a factor of \FlowLongU. These figures describe flows, since voluntary long reviews could have risen more without the prompt.

\input{generated/tab_flow}

\subsection{Order: small changes near launch, erosion later}

With the change in the stream established, we begin with the task of choosing among products. H1 holds that a prompt raises the level of ratings and compresses their spread, while agreement with critics stays within a small margin near launch. At modest review counts, the data support this prediction (Table~\ref{tab:order}). Holding sampled review counts at 50 per game, the rank correlation with Metacritic across \McGames games is \McB before launch and \McC after, a change of \McDiff (95\% CI \McDiffLo to \McDiffHi). The lower bound of the 90\% interval (\McDiffLoNinety) lies inside the margin of \Margin, and OpenCritic gives the same answer across \OcGames games: \OcB and \OcC, a change of \OcDiff (90\% lower bound \OcDiffLoNinety).

The result depends on the number of reviews compared. In the smaller set of games with 200 reviews, the change is \McLDiff (95\% CI \McLDiffLo to \McLDiffHi), and its 90\% lower bound of \McLDiffLoNinety falls outside the margin. Noninferiority therefore holds at 50 reviews per game and is not established at 200. With every review in the store-wide dataset counted, the change is \NatDiff (95\% CI \NatLo to \NatHi) across \NatG games. The post-launch window of that dataset is short, with a median of \NatNPost reviews per game against \NatNPre before, so this comparison does not show a gain from volume.

\input{generated/tab_order}

\begin{figure}[H]
\centering
\includegraphics[width=\textwidth]{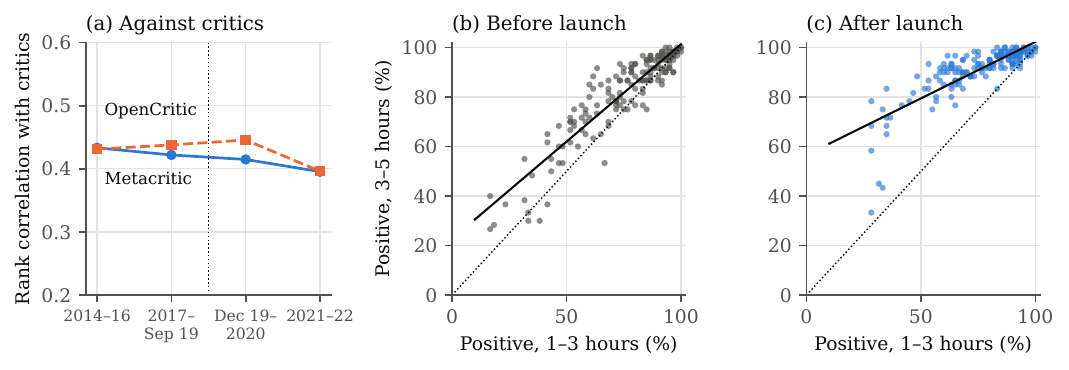}
\caption{Order. Panel (a): rank correlation with critics, holding sampled reviews at 50 per game. Panels (b) and (c): each point is a game, comparing players at 1--3 hours with players at 3--5 hours in one draw of 60 reviews per band. The dotted line marks equality and the solid line is the fitted slope.}
\label{fig:order}
\end{figure}

Later years look different, and we report them separately because we have no design that attributes the later change to the prompt. By 2021--22, the correlation at 50 reviews is \McD, a change of \McDiffD from the pre-period (95\% CI \McDiffDLo to \McDiffDHi). At 200 reviews it is \McLD, a change of \McLDiffD (95\% CI \McLDiffDLo to \McLDiffDHi). Because critic scores are mostly fixed near release, they benchmark a persistent ordering, so falling agreement can reflect games that changed as well as a rating that became less accurate.

Adjacent bands lead to the same reading. Across \VGames games, with 60 reviews per band, the 3--5 hour band correlates with Metacritic at \VPreA before launch and \VPostA after, and the 1--3 hour band at \VPreB and \VPostB, for a difference in differences of \VDid (95\% CI \VDidLo to \VDidHi; Table~\ref{tab:orderdetail}, Panel B). Level and spread, by contrast, change as H1 predicts. The spread of the 3--5 hour band narrows from \VSdPreA to \VSdPostA, with less change in the 1--3 hour band (\VSdPreB to \VSdPostB), and Figure~\ref{fig:order}(b, c) shows scores lifting and compressing toward the ceiling, with the slope falling from \VSlPre to \VSlPost.

\input{generated/tab_order_detail}

Proposition~1's common multiplier is a useful benchmark but only an approximation. Across \MultG games with enough reviews, the prompt multiplies the odds of a positive review by about \MultOdds on average, and the standard deviation of the log multiplier net of sampling noise is \MultSD. Lower-scoring games gain more (rank correlation with the pre-launch score of \MultCorrScore), which compresses the range. Rankings also shift somewhat: the rank correlation between the two bands falls from \VCoPre to \VCoPost, and game rankings persist at \StabLa across the launch, against \StabPl across an earlier boundary (Table~\ref{tab:orderdetail}, Panel A). The ordering therefore holds against critics, although it is not untouched.

New titles show one descriptive gain. A game's first 30 days predict its score in months 6--18 at a rank correlation of \NewPost for titles released after launch, against \NewPre before, and the mean gap between the early and later score falls from \NewGapPre to \NewGapPost points (Table~\ref{tab:orderdetail}, Panel C). Because release cohorts differ and later scores are themselves exposed to the prompt, this comparison does not identify an effect on early accuracy.

\subsection{Responsiveness: the score moves far less above the threshold}

\begin{figure}[H]
\centering
\includegraphics[width=\textwidth]{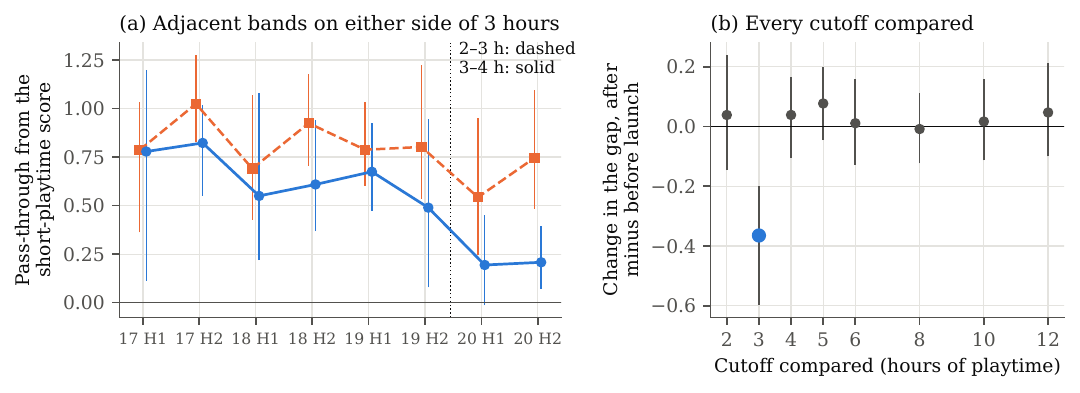}
\caption{Responsiveness. Panel (a): pass-through from the same game's short-playtime score by half-year, with 95\% bootstrap intervals; the dotted line marks the launch. Panel (b): change across the launch in the pass-through gap between adjacent one-hour bands on either side of each cutoff, with 95\% bootstrap intervals.}
\label{fig:responsiveness}
\end{figure}

The task of noticing change gives a sharply different picture. H2a and H2b agree that percent pass-through falls above the threshold relative to below it, and it does (Tables~\ref{tab:pass} and~\ref{tab:gap}, Figure~\ref{fig:responsiveness}). Before launch, the adjacent bands behave alike, with pass-through of \PTbPre at 2--3 hours and \PTaPre at 3--4 hours. After launch, it falls to \PTaPost at 3--4 hours and to \PTbPost at 2--3 hours. Further above the threshold it falls by similar proportions, for example from \PTmPre to \PTmPost at 10--30 hours, a band with prompts of its own. Excluding the sale months (June, July, November and December), when counts and positivity spike at every playtime, sharpens the contrast: the 2--3 hour band barely moves (\NSbPre to \NSbPost), while the 3--4 hour band falls from \NSaPre to \NSaPost.

\input{generated/tab_pass}

The pass-through gap sharpens the test by comparing the two bands in the same game and month. The gap is \GapPre before launch and \GapPost after, a change of \GapDiff (SE \GapDiffSE; 95\% CI \GapDiffLo to \GapDiffHi). First stages are strong, with $F$ statistics of \GapFPre before launch and \GapFPost after, and the change ranges from \SeedMin to \SeedMax across 20 split seeds. It is \NSGapDiff (SE \NSGapDiffSE) excluding sale months and \FRGapDiff (SE \FRGapDiffSE) when measured against pre-launch months under Valve's filter.

The break appears where and when the prompt does. Pairs of bands entirely on one side of the threshold show no change, at \PlBelow (SE \PlBelowSE) below it and at \PlAbove (SE \PlAboveSE) and \PlAboveB (SE \PlAboveBSE) above it. Figure~\ref{fig:responsiveness}(b) repeats the comparison for cutoffs from 2 to 12 hours, with a count rule of 20 short-playtime reviews so that all cutoffs can be compared: the change is \CutMain at three hours and between \CutOthLo and \CutOthHi at all others. Over time, the gap is \PPa in 2017, \PPb in 2018 and \PPc in 2019, and it then moves to \PPd for December 2019 to June 2020 and \PPe after that. Fifteen placebo launch dates within the pre-period give changes between \FakeMin and \FakeMax, none as large as the true one.

\input{generated/tab_gap}

The log-odds scale separates the two hypotheses, because the ceiling acts only on the percent scale. The 2--3 hour band does not change (\LObPre before launch and \LObPost after), whereas the 3--4 hour band falls from \LOaPre to \LOaPost, so the gap changes by \GapLoDiff (SE \GapLoDiffSE; 95\% CI \GapLoDiffLo to \GapLoDiffHi). This result supports H2b over H2a, with less precision than on the percent scale: part of the loss goes beyond the arithmetic of the ceiling. Because small cells can distort log-odds estimates, we applied the estimator with its continuity correction to simulated counts with each period's actual cell sizes and a known slope. The bias is at most \SimBias in absolute value and is similar in the two periods (Appendix~\ref{app:sim}).

The reference signal also weakens, which matters for what follows. Net of binomial sampling noise, within-game movement falls from \SDaPre to \SDaPost points in the 3--4 hour band and from \SDbPre to \SDbPost in the 2--3 hour band, and the short-playtime score itself moves less, from \SDsPre to \SDsPost points. Its split-half reliability within game and month falls from \RelPre to \RelPost at the same median count of \RelN reviews. Because that fall affects every band, a pre/post fall in detection against this score cannot by itself be attributed to the prompt.

Equation~\eqref{eq:ratio} shows how much of the percent loss comes from scale. For a deterministic relation between target score $R$ and reference score $U$,
\begin{equation}
\frac{dR}{dU} = \frac{R(1-R)}{U(1-U)} \times \frac{d\,\mathrm{logit}(R)}{d\,\mathrm{logit}(U)},
\label{eq:decomp}
\end{equation}
so a cross-period comparison needs the reference's scale as well as the target's. On the paired sample (Table~\ref{tab:theory}), the target's ceiling factor is \DecTarget, the reference's factor is \DecRef, and the log-odds factor is \DecLo, with a product of \DecProd against an observed percent ratio of \DecObs. The match is approximate, because regression weights and heterogeneity enter. As an illustration, starting from the model's pre-launch value of \DecPrePred, the ceiling alone takes percent pass-through to \DecNoSensor, the log-odds fall alone takes it to \DecNoCeiling, and the two together give \DecPostPred, close to the estimated \DecPostObs. Ceiling and response thus contribute about equally.

The sensor factor can also be checked against its components. In the 3--4 hour band, added reviews make up \AddA\% of praise and \AddB\% of complaints. Before launch, a move in the short-playtime score runs \CarryW\% through praise and the rest through complaints (95\% CI \CarryWLo\% to \CarryWHi\%). If added reviews did not respond at all, Equation~\eqref{eq:phi} would give $\phi =$ \PhiNone. The observed log-odds ratio is \PhiObs, or \PhiRel relative to the band below (95\% CI \PhiRelLo to \PhiRelHi), which implies $\eta \approx$ \EtaImp (95\% CI \EtaImpLo to \EtaImpHi). On the paired sample the relative ratio is \PhiPair and $\eta \approx$ \EtaPair (95\% CI \EtaPairLo to \EtaPairHi). The strong form of the voice account ($\eta = 0$) fails on both samples, so added reviews do respond. Selection theory's silent-majority benchmark ($\eta = 1$) fails on the paired sample and is not excluded on the band-specific samples. Taken together with the separate calculation in Section~\ref{sec:added}, the estimates center on a response about half that of voluntary reviews.

\input{generated/tab_theory}

\subsection{Sensitivity: what the stream still knows}

\begin{figure}[H]
\centering
\includegraphics[width=\textwidth]{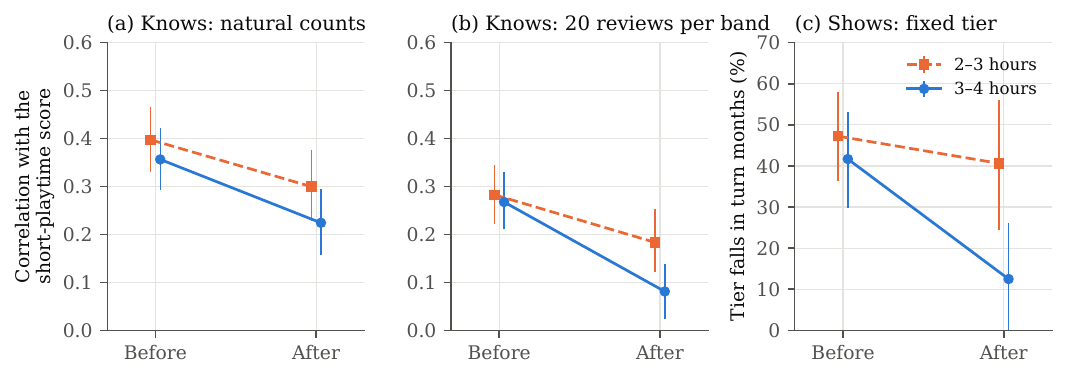}
\caption{What the stream knows and what the display shows, on the same game-months. Panels (a) and (b): correlation of each band's log odds with the short-playtime score, within game and month, at natural counts and at 20 reviews per band. Panel (c): how often a fixed tier falls in a turn month. Bars show 95\% bootstrap intervals.}
\label{fig:knows}
\end{figure}

A score that moves less need not know less, so we turn to how well the stream can still separate change from noise. H3a and H3b disagree about natural counts (Table~\ref{tab:knows}, Figure~\ref{fig:knows}). At natural counts, on the \CorrNatN paired game-months, the correlation of the 3--4 hour band with the short-playtime score falls from \CorrNataPre to \CorrNataPost and that of the 2--3 hour band from \CorrNatbPre to \CorrNatbPost, so the threshold-specific change is \CorrNatDid (95\% CI \CorrNatLo to \CorrNatHi).

Equal counts tell a different story. With 20 reviews per band, the correlation falls from \CorrFixaPre to \CorrFixaPost above the threshold and from \CorrFixbPre to \CorrFixbPost below it, for a threshold-specific change of \CorrFixDid (95\% CI \CorrFixLo to \CorrFixHi). This interval just reaches zero. At 40 reviews per band the change is \CorrFortyDid (95\% CI \CorrFortyLo to \CorrFortyHi), and on the percent scale at 20 reviews it is \CorrFixPctDid (95\% CI \CorrFixPctLo to \CorrFixPctHi). These results support H3b over H3a. Per review, the stream above the threshold carries less news, but as a stream it has not detectably lost ground relative to the stream below, because more reviews lower its noise.

\input{generated/tab_knows}

Equation~\eqref{eq:detect} accounts for this pattern. With the log-odds ratio of \PhiRel relative to the band below, and the scores and counts in each band, it gives a relative sensitivity of \PredFix at equal counts and \PredNat at natural counts (Table~\ref{tab:theory}, Panel C). The observed ratios are \ObsFix (95\% CI \ObsFixLo to \ObsFixHi) and \ObsNat (95\% CI \ObsNatLo to \ObsNatHi). At equal counts, where sampling noise dominates, the match is close. At natural counts the model gives an upper limit, because added reviews can reduce only the sampling part of a band's noise.

Rank separation of turn and calm months agrees, with less precision. AUC falls in both bands, from \AUCaPre to \AUCaPost above the threshold and from \AUCbPre to \AUCbPost below it, and the threshold-specific change is \AUCDid (95\% CI \AUCLo to \AUCHi). Expressed as the sensitivity index $d_a$, the fall is from \DAaPre to \DAaPost above and from \DAbPre to \DAbPost below, a threshold-specific change of \DADid (95\% CI \DALo to \DAHi). A fall in both bands is what the lower reliability of the short-playtime score would produce, and the contrast removes it. The data limit any threshold-specific fall in correlation across all months to \CorrNatLoAbs, whereas the tail of the distribution is estimated less precisely: in cells with at least 40 reviews per band, the hit rate at a 10\% false-alarm rate changes by \BigTTen points relative to the band below (95\% CI \BigTTenLo to \BigTTenHi), against \MainTTen points in the full sample.

\subsection{Display: what the rating shows}
\label{sec:display}

What a rating shows is a separate question, and H4 addresses it. We begin with adjacent bands, where the comparison is cleanest (Table~\ref{tab:display}, Figure~\ref{fig:knows}c). On common game-months, a fixed tier falls in \DHaPre\% of turn months just above the threshold before launch and \DHaPost\% after, whereas just below the threshold the rate goes from \DHbPre\% to \DHbPost\%. The threshold-specific change is \DHDid points (95\% CI \DHDidLo to \DHDidHi). False alarms in calm months fall by similar amounts in both bands (\DFaPre\% to \DFaPost\% above and \DFbPre\% to \DFbPost\% below), with a threshold-specific change of \DFDid points. Relative to the band below, a fixed tier above the threshold therefore loses hits without a matching fall in false alarms. These rates rest on \DTnPre turn months before launch and \DTnPost after.

Detection measures make the same point. The tier's criterion rises from \CRaPre to \CRaPost above the threshold and from \CRbPre to \CRbPost below it, a threshold-specific change of \CRDid (95\% CI \CRLo to \CRHi). Its sensitivity falls as well, from \DPaPre to \DPaPost above the threshold, while it rises from \DPbPre to \DPbPost below it (threshold-specific change \DPDid, 95\% CI \DPLo to \DPHi). At this level, then, the coarse tier turns conservative and its point estimates also suggest a loss of information, whereas the statistic computed from the same reviews shows no threshold-specific loss of sensitivity.

\input{generated/tab_display}

A threshold set for the new scale does better. We compare the tier with a log-odds fall, using thresholds from other games chosen to match the tier's false-alarm rate. Before launch the two rules perform about equally in both bands (gains of \GainaPre and \GainbPre points). After launch, the recalibrated rule catches \GainaPost points more than the tier above the threshold and \GainbPostAbs points fewer below it, a post-launch contrast of \GainCon points (95\% CI \GainConLo to \GainConHi). Net of the small pre-launch gains, the threshold-specific change is \GainDid points (95\% CI \GainDidLo to \GainDidHi). At fixed false-alarm rates of 5, 10 and 20\%, the threshold-specific change in the hit rate is \TFiveDid, \TTenDid and \TTwentyDid, all with wide intervals around zero. This evidence fits a display that has gone quiet without a detectable loss in the stream, which is what H4 predicts.

The reconstructed public score brings the analysis closer to what a consumer sees. It has no adjacent control, so it reflects the general fall in signal as well as the prompt (Table~\ref{tab:public}). Because display outcomes need three prior months in a regime, they run from \DatePreOne to \DatePreOneEnd and from \DatePreTwo to \DatePreTwoEnd before launch, and from \DatePost to \DatePostEnd after it. Estimates average over \PuSplits random halves of short-playtime reviews. In turn months, the 30-day score falls \PuDPPreAbs points before launch and \PuDPPostAbs after, and its label falls in \PuHPre\% and \PuHPost\% of cases. In calm months, the label falls in \PuFPre\% and \PuFPost\%. Hits and false alarms fall by similar proportions, so the criterion rises from \PuCrPre to \PuCrPost, while $d'$ goes from \PuDprimePre to \PuDprimePost.

The downturns themselves differ in severity across periods, since the short-playtime fall averages \PuDUPreAbs points before launch and \PuDUPostAbs after. For the diagnostic score above three hours, whose reference is the score under three hours, we can compare periods after standardizing to a common distribution of falls. The fall is \SevPreAbs points in the pre-launch months under the filter and \SevPostAbs after, a difference of \SevDiff points (95\% CI \SevLo to \SevHi), and the tier falls in \SevDownPre\% and \SevDownPost\% of cases. Table~\ref{tab:label} in the Appendix gives the unstandardized rows.

Removing the added reviews while holding the post-launch months fixed shows the prompt's effect on the label more directly. We use shares estimated from same-game flows and bracket the response of added reviews with two cases. Removing them as a fixed monthly mass treats them as unresponsive ($\eta = 0$): the score falls \CfFixAbs points in turn months and the label falls in \CfFixHit\% of them, at a false-alarm rate of \CfFixFa\%. Removing them in proportion treats them as fully responsive ($\eta = 1$): the figures are \CfPropAbs points, \CfPropHit\% and \CfPropFa\%. The actual values are \PuDPPostAbs points, \PuHPost\% and \PuFPost\%, so added reviews lower hits and false alarms together. Against the fixed-mass case, the change in hits is \CfHitEst points (95\% CI \CfHitLo to \CfHitHi) and in false alarms \CfFaEst points (95\% CI \CfFaLo to \CfFaHi). Against the proportional case, these changes are \CfPropHitEst points (95\% CI \CfPropHitLo to \CfPropHitHi) and \CfPropFaEst points (95\% CI \CfPropFaLo to \CfPropFaHi). Because this comparison holds the months and the short-playtime signal fixed, it isolates the added reviews. They raise the label's criterion by \CfCrDiff against the fixed-mass case (95\% CI \CfCrLo to \CfCrHi) and by \CfPropCrDiff against the proportional case (95\% CI \CfPropCrLo to \CfPropCrHi), and a response of about one half suggests an effect between the two. They do not lower its sensitivity in either case (\CfDpDiff, 95\% CI \CfDpLo to \CfDpHi; \CfPropDpDiff, 95\% CI \CfPropDpLo to \CfPropDpHi). On the public score, then, the evidence points to a shift in the operating point and not to a loss of sensitivity.

It follows that recalibrating the public score is a choice of operating point. On held-out games after launch, the label's \PuHPost\% hit rate at a \PuFPost\% false-alarm rate is as good as other rules achieve at that rate: \DetPctM\% for a fall in points and \DetLogM\% for a fall in log odds. A complaint share relative to its norm, a natural candidate for an alert, does no better than a plain fall in points at a false-alarm rate of 10\%: \DetShareTen\% against \DetPctTen\%, a difference with an interval of \DetShareLo to \DetShareHi points. A complaint count relative to its norm catches \DetCountTen\%. Accepting more false alarms does buy more hits, since a fall in points catches \DetPctTen\% at a 10\% false-alarm rate and \DetPctTwenty\% at 20\%, against \DetPreTen\% and \DetPreTwenty\% before launch. These hit rates rest on about \PuNT turn months after launch. Thresholds learned before launch do not transfer: a five-point fall flagged \StPreHit\% of turn months and \StPreFa\% of calm months before launch, and \StPostHit\% and \StPostFa\% after.

\input{generated/tab_public}

\subsection{Channel: which changes stay visible}
\label{sec:channel}

So far, we have measured change against short-playtime reviews. H5 turns to events coded from developers' own words, and the first finding concerns short-playtime reviews themselves. Relative to quiet windows, the short-playtime score does not fall at documented failures: the change is \FaSPre points before launch (SE \FaSPreSE) and \FaSPost after (SE \FaSPostSE; Table~\ref{tab:failures}), and its rank separation of failure windows from quiet windows is \FaSAUCPre and \FaSAUCPost. The empirical claims above therefore concern short-playtime evaluations and not product change in general. These players are the group most like a new buyer, and their score may matter for that reason, but it is not a general sensor of failures.

\input{generated/tab_failures}

\begin{figure}[H]
\centering
\includegraphics[width=\textwidth]{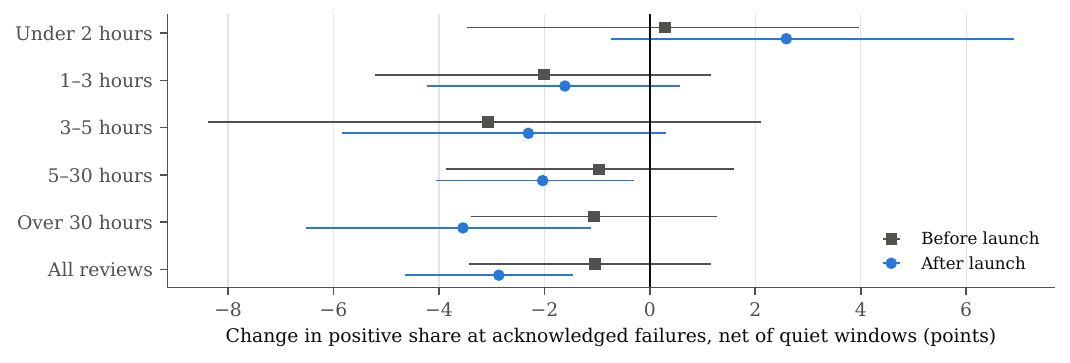}
\caption{Documented failures. Change in each band's positive share in failure windows, net of quiet windows in the same games, before and after launch. Bars show 95\% bootstrap intervals.}
\label{fig:failures}
\end{figure}

The score over all reviews does register failures, and not less so after launch. Net of quiet windows, it falls \FaAllPreAbs points before launch (SE \FaAllPreSE) and \FaAllPostAbs after (SE \FaAllPostSE), a change of \FaAllDid (95\% CI \FaAllLo to \FaAllHi). The log-odds response is \FaAllLoPre and \FaAllLoPost, and rank separation rises from \FaAllAUCPre to \FaAllAUCPost. The estimate rests on \FaEpPost usable post-launch failures (\FaMajPost major), and its interval limits any fall in the response to \FaAllHi points. Players over 30 hours respond most after launch (\FaHPost points, SE \FaHPostSE; Figure~\ref{fig:failures}), and the reconstructed public score behaves the same way, with a change net of quiet windows of \PFdpPre points before launch and \PFdpPost after.

These failures run through complaints. Net of quiet windows, negative reviews per day rise \FaNegPre log points before launch and \FaNegPost after, on a complaint base that the prompt has made larger, while positive reviews move by \FaPosPre and \FaPosPost. This pattern is close to the boundary case in Proposition~2. Complaints are less inflated than praise (\AddB\% against \AddA\% added in the 3--4 hour band), and they grow in the same proportion on a larger base after launch, which suggests that prompted players report failures when they meet them. In the threshold gap, the difference between 3--5 hours and 1--3 hours at failures changes by \FaGapDid points across the launch (95\% CI \FaGapLo to \FaGapHi). Together, these results support H5.

Surges in negative reviews point the same way. We call a month a surge when its negative reviews reach three times their trailing median and at least 100. Such months cost the score \SurgeAAbs points (SE \SurgeASE) from 2017 to February 2019 (\SurgeAn months), \SurgeBAbs points (SE \SurgeBSE) from June to September 2019 (\SurgeBn months) and \SurgePAbs points (SE \SurgePSE) after launch (\SurgePn months). The cost falls with the arrival of Valve's filter in March 2019, before the prompt, and the score still falls at surges after the prompt.

Dated updates are a broader and more mixed class of events (Table~\ref{tab:updates}, Figure~\ref{fig:updates} in the Appendix). Before launch, both bands respond alike: log odds change by \UMbPre at 1--3 hours and \UMaPre at 3--5 hours, a gap of \UMgPre (SE \UMgPreSE). After launch, the 1--3 hour band falls by \UMbPostAbs and the 3--5 hour band by \UMaPostAbs, a gap of \UMgPost (SE \UMgPostSE). The gap changes by \UMch (SE \UMchSE), against \QMch (SE \QMchSE) in quiet windows, which leaves a difference of \UMddd (95\% CI \UMdddLo to \UMdddHi), and isolated updates give \UIso (95\% CI \UIsoLo to \UIsoHi). The sign fits a band that responds less, and the estimates are suggestive. Pooled percent changes, which weight windows by review counts and so favor large surges in reviewing, show no threshold-specific change (\UPct, 95\% CI \UPctLo to \UPctHi). Allowing each update its own direction, the slope of the 3--5 hour change on the 1--3 hour change falls from \UIVPre to \UIVPost (change 95\% CI \UIVLo to \UIVHi).

\input{generated/tab_updates}

\subsection{Out-of-sample check: the other prompts}

\begin{figure}[H]
\centering
\includegraphics[width=\textwidth]{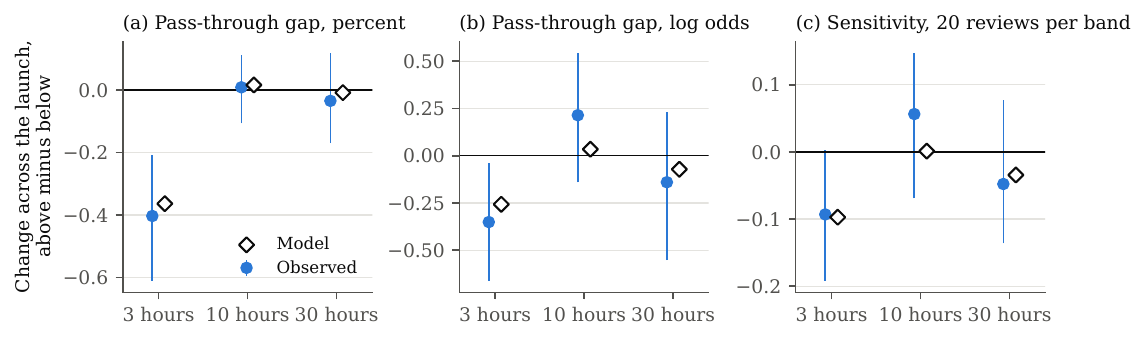}
\caption{The model at all three prompts. Each panel shows the change across the launch in adjacent bands on either side of the prompt, above minus below. Bars show 95\% bootstrap intervals. The three-hour values calibrate the model; the 10-hour and 30-hour values are out of sample.}
\label{fig:thresholds}
\end{figure}

We built the model on the three-hour prompt, so the prompts at 10 and 30 hours offer a check it has not seen (Table~\ref{tab:thresholds}, Figure~\ref{fig:thresholds}). Added reviews make up far less of the stream at those prompts: \OtATen and \OtAThirty of praise, against \OtAThree at three hours, and \OtBTen and \OtBThirty of complaints, against \OtBThree. The model therefore implies breaks close to zero. The observed percent gap changes by \OtObsTen at 10 hours (95\% CI \OtLoTen to \OtHiTen) and \OtObsThirty at 30 hours (95\% CI \OtLoThirty to \OtHiThirty), against model values of \OtPredTen and \OtPredThirty, and against \OtObsThree at three hours, where the model gives \OtPredThree. Both intervals exclude a break of the three-hour size, and every model value in Table~\ref{tab:thresholds} lies inside the observed interval. The log-odds and sensitivity rows have wide intervals, so they rule out large departures without confirming the model's exact magnitudes. The pattern fits a dose response: the larger the added share, the larger the threshold-specific break.

\input{generated/tab_thresholds}

\subsection{Added reviews: what they do and do not carry}
\label{sec:added}

The model takes the behavior of added reviews as an input, so we now examine them directly. Three findings bear on the mechanisms behind H2 and H3 (Table~\ref{tab:mech}).

First, added reviews respond about half as much as voluntary ones in log odds. After launch, in each game's worst fifth of months by the short-playtime score, reviews at 2--3 hours are \WorstOrg\% positive against \OtherOrg\% in other months, a gap of \WorstOrgGap points (SE \WorstOrgGapSE). Added reviews just above the threshold are \WorstExtra\% positive against \OtherExtra\%, a gap of \WorstExtraGap points (SE \WorstExtraGapSE). In percentage points, the ceiling hides most of this difference. In log odds, the added reviews fall \WorstLoExtra and the band below \WorstLoOrg, a ratio of \WorstEta (95\% CI \WorstEtaLo to \WorstEtaHi) that is close to the value of \EtaImp implied by pass-through. Before launch, the 3--4 hour and 2--3 hour bands fall together in such months (\WorstPreA and \WorstPreB points).

Second, added reviews mention updates, developers and breakage less, although most of the gap comes from length. Among English reviews, the share containing these words is \CWExtra\% among added reviews against \CWBelow\% just below. Shorter reviews offer fewer chances to use a word, and added reviews are short: \CWShortAdded\% of them have under 50 characters, against \CWShortBelow\% of reviews just below. If added reviews had the below-threshold rate within each length bin, their share would be \CWStd\%, and within the longest bin the rate is \CWLongAdded\% against \CWLongBelow\%. Because length may itself respond to the prompt, this decomposition is descriptive. The word rule is also a loose measure. In a coding of \KwN sampled reviews, blind to the rule, only \KwPrec\% of those the rule flags, and \KwUn\% of those it does not, report that the game has changed, judged on the first 210 characters. We therefore treat the rule as a measure of mentions and not of reports of change.

Third, longer reviews stay more responsive. We read this as heterogeneity and not as a test of prompted status, since prompts can produce long reviews and voluntary reviews can be short. Among reviews above three hours, pass-through for reviews of 200 or more characters goes from \PTlgPre to \PTlgPost (log odds \LOlgPre to \LOlgPost), and for shorter reviews from \PTshPre to \PTshPost (log odds \LOshPre to \LOshPost).

The interface offers a rival account of the lower response: a prompted reviewer need not see the score, so the fall in pass-through could reflect a lost echo of the displayed score and not a weaker signal. The data give little support to this account. We take the prior month's positive share among Steam purchases as the score a reviewer could have seen, on the \VisN paired game-months where it is available. Holding that score fixed leaves the change in the pass-through gap nearly unchanged: \VisPctBase without it and \VisPctCtrl with it on the percent scale (95\% CI \VisPctCtrlLo to \VisPctCtrlHi), and \VisLoBase and \VisLoCtrl on the log-odds scale (95\% CI \VisLoCtrlLo to \VisLoCtrlHi). Holding the lifetime score fixed instead gives \VisLifePct and \VisLifeLo. A band's own loading on the prior month's score does not fall more above the threshold either: it goes from \VisRhoAPre to \VisRhoAPost above and from \VisRhoBPre to \VisRhoBPost below, a threshold-specific change of \VisRhoDid (95\% CI \VisRhoLo to \VisRhoHi).

Added reviews do carry information about fit. After launch, a positive recommendation predicts further play about equally on either side of the threshold: \TPa (SE \TPaSE) above and \TPb (SE \TPbSE) below. This result supports their relevance to a person's own experience of a game, although it does not speak to their accuracy about change over time.

\input{generated/tab_mech}

\subsection{Support and robustness}

We close the empirical analysis by examining how firm these estimates are. Table~\ref{tab:robust} re-estimates the pass-through gap under alternative samples, with the full estimator and bootstrap in each row. On the percent scale, the change ranges from \VarLo to \VarHi, keeps its sign in every row, and has an interval that excludes zero in \VarSigN of \VarN rows. On the log-odds scale, it ranges from \VarLoLo to \VarLoHi and keeps its sign, but its interval excludes zero in only \VarLoSigN rows. The percent result is therefore firm and the log-odds result weaker: among the \BothG games observed in both periods, the log-odds change is \BothLo (95\% CI \BothLoLo to \BothLoHi), against \BothDiff (95\% CI \BothCILo to \BothCIHi) on the percent scale.

Comparable support changes magnitudes without changing the pattern. With equal windows of 14 months on each side, the change is \EqDiff (SE \EqSE). In games at least a year past their first review, it is \AgeDiff (SE \AgeSE), which suggests that part of the baseline gap reflects young games. Dropping the games with removed periods gives \HoleDiff (SE \HoleSE). The percent result also holds for Steam purchases only, for English reviews, and for reviews never edited, with a one-hour tolerance or none, which limits the role of Valve's related prompt to revise a review.

The composition of short-playtime reviews also shifts across the launch. Steam purchases fall from \RefBuyPre\% to \RefBuyPost\% of these reviews, the median reviewer owns \RefGamesPost games against \RefGamesPre before, and the median game is older, at \RefAgePost days past its first review against \RefAgePre (Table~\ref{tab:summary}). Limiting short-playtime reviews to English Steam purchases gives \EngDiff with a wide interval (95\% CI \EngLo to \EngHi; first-stage $F$ of \EngF).

Finally, month effects absorb shocks common to all games, but not every pandemic effect specific to a game or a playtime. Using only December 2019 to February 2020 gives \PanDiff with a very wide interval (95\% CI \PanLo to \PanHi, \PanN game-months, first-stage $F$ of \PanF), so this window offers limited evidence on that question.

\input{generated/tab_robust}

\subsection{Summary of hypothesis tests}
\label{sec:summary}

Table~\ref{tab:hypotheses} collects the evidence. No single theory accounts for the pattern, and each is right about part of it. Volume lowers noise, as selection theory implies, but added reviews do not respond like voluntary ones. Added reviews do respond less, as voice theory implies, but the stream as a whole shows no detectable threshold-specific loss of sensitivity. Scores do compress, as inflation theory implies, but compression mainly moves the display's operating point. We now turn to what this pattern means for theory and for the design of rating systems.

\input{generated/tab_hypotheses}

%% file: generated/tab_prompt.tex
\begin{table}[H]
\centering
\caption{What changes in the review stream at three hours of play}
\label{tab:prompt}
\footnotesize\setlength{\tabcolsep}{4pt}
\begin{tabular}{lrrrrrrr}
\toprule
Outcome & Mean below & \multicolumn{2}{c}{Jump before launch} & \multicolumn{2}{c}{Jump after launch} & \multicolumn{2}{c}{Change}  \\
\midrule
Positive review (pp) & 85.9 & +0.8 & (0.3) & +7.4 & (0.7) & +6.6 & (0.8) \\
Under 20 characters (pp) & 32.8 & +0.5 & (0.4) & +14.8 & (0.5) & +14.4 & (0.7) \\
Log review length & 3.79 & $-$0.03 & (0.01) & $-$0.53 & (0.02) & $-$0.50 & (0.02) \\
One review on record (pp) & 16.5 & +0.5 & (0.3) & +3.5 & (0.5) & +3.0 & (0.6) \\
Log games owned & 3.89 & +0.01 & (0.01) & $-$0.40 & (0.03) & $-$0.41 & (0.03) \\
Any helpful vote (pp) & 25.4 & $-$0.3 & (0.4) & $-$8.1 & (0.5) & $-$7.8 & (0.7) \\
Played 1+ hour after reviewing (pp) & 56.5 & +1.6 & (0.5) & +13.6 & (0.8) & +11.9 & (0.9) \\
Update, developer or breakage words, English (pp) & 5.8 & $-$0.1 & (0.4) & $-$2.7 & (0.4) & $-$2.6 & (0.5) \\
\bottomrule
\end{tabular}
\par\vspace{2pt}{\footnotesize\raggedright Local linear jump at 180 minutes, bandwidth 30 minutes, game fixed effects. The two periods are estimated jointly; standard errors clustered by game in parentheses. ``Mean below'' is the post-launch mean at 150--179 minutes.\par}
\end{table}

%% file: generated/tab_flow.tex
\begin{table}[H]
\centering
\caption{Praise and complaints per game per day, same games}
\label{tab:flow}
\small
\begin{tabular}{lrrrrrrr}
\toprule
Playtime at review & \multicolumn{3}{c}{Positive reviews} & \multicolumn{3}{c}{Negative reviews} & Added reviews \\ & Before & After & Ratio & Before & After & Ratio & positive (\%) \\
\midrule
Under 3 hours & 2.11 & 2.49 & 1.18 & 0.94 & 0.77 & 0.81 & -- \\
3--10 hours & 3.51 & 14.64 & 4.17 & 0.65 & 1.01 & 1.57 & 97 \\
10--30 hours & 5.10 & 15.58 & 3.05 & 0.82 & 1.03 & 1.26 & 98 \\
Over 30 hours & 12.65 & 32.95 & 2.60 & 2.75 & 3.63 & 1.32 & 96 \\
\midrule
All playtimes & 23.37 & 65.66 & 2.81 & 5.15 & 6.44 & 1.25 &  \\
\bottomrule
\end{tabular}
\par\vspace{2pt}{\footnotesize\raggedright 271 games with at least 300 reviews in each period. Before is January 2018--September 2019; after is December 2019--December 2020. The last column is the positive share of the increase in that band.\par}
\end{table}

%% file: generated/tab_order.tex
\begin{table}[H]
\centering
\caption{Order: rank correlation between player scores and critics}
\label{tab:order}
\scriptsize\setlength{\tabcolsep}{3pt}
\begin{tabular}{lrrrrrlrc}
\toprule
Benchmark & Games & 2014--16 & 2017--Sep 19 & Dec 19--2020 & 2021--22 & Launch change [95\% CI] & 90\% lower & Within \Margin \\
\midrule
Metacritic, 50 per game & 1,066 & 0.43 & 0.42 & 0.41 & 0.40 & $-$0.007 [$-$0.034, 0.016] & $-$0.029 & Yes \\
Metacritic, 200 per game & 646 & 0.43 & 0.42 & 0.39 & 0.35 & $-$0.032 [$-$0.065, 0.000] & $-$0.060 & No \\
OpenCritic, 50 per game & 623 & 0.43 & 0.44 & 0.45 & 0.40 & 0.008 [$-$0.031, 0.046] & $-$0.023 & Yes \\
OpenCritic, 200 per game & 370 & 0.44 & 0.46 & 0.44 & 0.37 & $-$0.023 [$-$0.078, 0.030] & $-$0.071 & No \\
\midrule
Metacritic, natural volume & 2,342 & & 0.49 & 0.49 & & $-$0.005 [$-$0.028, 0.018] & & \\
OpenCritic, natural volume & 2,047 & & 0.47 & 0.50 & & 0.026 [$-$0.003, 0.054] & & \\
\bottomrule
\end{tabular}
\par\vspace{2pt}{\footnotesize\raggedright Spearman correlation across games. The upper rows hold sampled reviews constant: each game contributes the same number of randomly drawn reviews to every period, and the correlation is computed within each of 40 draws and averaged. Natural-volume rows use every review in the store-wide data. Launch change compares December 2019--2020 with 2017--September 2019; intervals bootstrap the same statistic over games. The last column asks whether the 90\% lower bound lies above $-$\Margin.\par}
\end{table}

%% file: generated/tab_order_detail.tex
\begin{table}[H]
\centering
\caption{Order: rank stability, adjacent bands, and early ratings}
\label{tab:orderdetail}
\small
\begin{tabular}{lrr}
\toprule
Measure & \multicolumn{2}{c}{Estimate} \\
\midrule
\multicolumn{3}{l}{\textit{Panel A. Rank order of games across two periods}} \\
Earlier boundary (2014--16 to 2017--Sep 19), 2,349 games & \multicolumn{2}{r}{0.83 [0.82, 0.84]} \\
Across the launch (2017--Sep 19 to Dec 19--2020), 2,954 games & \multicolumn{2}{r}{0.78 [0.77, 0.79]} \\
\midrule
\multicolumn{3}{l}{\textit{Panel B. Adjacent bands, 163 games, 60 reviews per band}} \\
 & Before launch & After launch \\
Correlation with Metacritic, 1--3 hours & 0.40 & 0.40 \\
Correlation with Metacritic, 3--5 hours & 0.35 & 0.37 \\
Standard deviation across games, 1--3 hours & 0.20 & 0.18 \\
Standard deviation across games, 3--5 hours & 0.17 & 0.10 \\
Slope of 3--5 hours on 1--3 hours & 0.77 & 0.47 \\
Rank correlation of the two bands & 0.89 & 0.80 \\
\midrule
\multicolumn{3}{l}{\textit{Panel C. New titles: first 30 days against months 6--18 (descriptive)}} \\
 & Jun 2016--Mar 2018 & Dec 2019--May 2021 \\
Rank correlation [95\% CI] & 0.71 [0.68, 0.74] & 0.74 [0.72, 0.77] \\
Mean absolute gap (pp) & 8.8 & 6.7 \\
Games & 1,532 & 1,793 \\
\bottomrule
\end{tabular}
\par\vspace{2pt}{\footnotesize\raggedright Panel A: rank correlation of game scores between two periods, 100 reviews per game and period, computed within each of 20 draws and averaged. Panel B: 60 reviews per game for each band and period, computed within each of 40 draws and averaged. Panel C: release dates by column, with the earlier cohort ending so that its months 6--18 precede the rollout; games with 10+ reviews in the first 30 days and 30+ in months 6--18, from the store-wide data. Brackets are 95\% intervals from a game bootstrap.\par}
\end{table}

%% file: generated/tab_pass.tex
\begin{table}[H]
\centering
\caption{Responsiveness: how much of the short-playtime score reaches each band}
\label{tab:pass}
\footnotesize
\begin{tabular}{lrrrrrrrrrr}
\toprule
Band & \multicolumn{5}{c}{Percent scale} & \multicolumn{5}{c}{Log-odds scale} \\ & \multicolumn{2}{c}{Before} & \multicolumn{2}{c}{After} & $F$ & \multicolumn{2}{c}{Before} & \multicolumn{2}{c}{After} & $F$ \\
\midrule
2--3 hours & 0.81 & (0.04) & 0.64 & (0.09) & 83 & 0.89 & (0.06) & 0.86 & (0.13) & 71 \\
3--4 hours & 0.73 & (0.05) & 0.18 & (0.05) & 66 & 0.85 & (0.07) & 0.57 & (0.10) & 54 \\
4--6 hours & 0.73 & (0.05) & 0.18 & (0.04) & 64 & 0.91 & (0.07) & 0.67 & (0.10) & 57 \\
6--10 hours & 0.65 & (0.05) & 0.13 & (0.04) & 60 & 0.85 & (0.07) & 0.47 & (0.09) & 53 \\
10--30 hours & 0.50 & (0.04) & 0.11 & (0.04) & 62 & 0.70 & (0.06) & 0.49 & (0.08) & 53 \\
Over 30 hours & 0.39 & (0.04) & 0.10 & (0.03) & 63 & 0.56 & (0.06) & 0.35 & (0.09) & 55 \\
\midrule
Over 3 hours, 200+ characters & 0.39 & (0.04) & 0.26 & (0.05) & 59 & 0.55 & (0.05) & 0.47 & (0.09) & 55 \\
Over 3 hours, under 200 characters & 0.49 & (0.04) & 0.10 & (0.02) & 64 & 0.77 & (0.06) & 0.51 & (0.07) & 52 \\
\bottomrule
\end{tabular}
\par\vspace{2pt}{\footnotesize\raggedright Coefficient of the band's monthly score on the same game's short-playtime score in the same month. Game and month fixed effects; repeated-measurement IV averaged over 40 random halves of the short-playtime reviews; standard errors from a bootstrap of whole game histories. $F$ is the first-stage statistic after launch. Game-months need 20+ reviews in the band and 30+ short-playtime reviews.\par}
\end{table}

%% file: generated/tab_gap.tex
\begin{table}[H]
\centering
\caption{Responsiveness: the pass-through gap at the threshold}
\label{tab:gap}
\footnotesize
\begin{tabular}{lrrrrrrl}
\toprule
Pair & \multicolumn{2}{c}{Before} & \multicolumn{2}{c}{After} & \multicolumn{3}{c}{Change [95\% CI]} \\
\midrule
\multicolumn{8}{l}{\textit{Panel A. Threshold pair: 3--4 hours minus 2--3 hours}} \\
Percent scale & $-$0.04 & (0.04) & $-$0.44 & (0.08) & $-$0.40 & (0.10) & [$-$0.57, $-$0.21] \\
Log-odds scale & $-$0.01 & (0.07) & $-$0.36 & (0.13) & $-$0.35 & (0.15) & [$-$0.65, $-$0.04] \\
Percent scale, excluding sale months & +0.00 & (0.06) & $-$0.53 & (0.14) & $-$0.53 & (0.16) & [$-$0.92, $-$0.29] \\
Log-odds scale, excluding sale months & $-$0.01 & (0.10) & $-$0.40 & (0.19) & $-$0.38 & (0.22) & [$-$0.90, $-$0.04] \\
Percent scale, pre-period Mar--Sep 2019 & +0.04 & (0.11) & $-$0.44 & (0.08) & $-$0.48 & (0.15) & [$-$0.79, $-$0.20] \\
Log-odds scale, pre-period Mar--Sep 2019 & $-$0.22 & (0.22) & $-$0.36 & (0.13) & $-$0.14 & (0.26) & [$-$0.66, 0.39] \\
\midrule
\multicolumn{8}{l}{\textit{Panel B. Placebo pairs, percent scale}} \\
2--3 hours minus 1--2 hours (both below) & $-$0.19 & (0.04) & $-$0.15 & (0.09) & +0.04 & (0.10) & [$-$0.15, 0.24] \\
4--6 hours minus 3--4 hours (both above) & $-$0.05 & (0.03) & +0.00 & (0.05) & +0.05 & (0.06) & [$-$0.07, 0.17] \\
6--10 hours minus 4--6 hours (both above) & $-$0.10 & (0.03) & $-$0.06 & (0.04) & +0.04 & (0.05) & [$-$0.04, 0.15] \\
\bottomrule
\end{tabular}
\par\vspace{2pt}{\footnotesize\raggedright Same specification as Table~\ref{tab:pass}, with the difference between the two bands in the same game-month as the outcome. 1,878 game-months across 222 games before launch and 1,241 across 167 after. The 2--3 hour minus 1--2 hour pair uses players under 1 hour as the short-playtime group. Sale months are June, July, November and December.\par}
\end{table}

%% file: generated/tab_theory.tex
\begin{table}[H]
\centering
\caption{Connecting the ratio model to the estimates}
\label{tab:theory}
\small
\begin{tabular}{lrr}
\toprule
Quantity & \multicolumn{2}{c}{Value} \\
\midrule
\multicolumn{3}{l}{\textit{Panel A. Percent pass-through and its parts, paired sample (after over before)}} \\
 & 3--4 hours & 2--3 hours \\
Target's ceiling factor, $R(1-R)$ & 0.53 & 0.95 \\
Reference's factor, inverse of $U(1-U)$ & 1.01 & 1.01 \\
Log-odds pass-through factor & 0.58 & 0.99 \\
Product & 0.31 & 0.95 \\
Observed percent pass-through ratio & 0.26 & 0.81 \\
\midrule
\multicolumn{3}{l}{\textit{Panel B. Sensor factor, Equation~(\ref{eq:phi}), 3--4 hours}} \\
Added share of praise, $a$ & \multicolumn{2}{r}{0.81} \\
Added share of complaints, $b$ & \multicolumn{2}{r}{0.57} \\
Part of a move carried by praise, $w$, pre-launch & \multicolumn{2}{r}{0.37 [0.19, 0.52]} \\
Predicted $\phi$ if added reviews do not respond & \multicolumn{2}{r}{0.34} \\
Observed log-odds ratio, relative to the band below & \multicolumn{2}{r}{0.70 [0.47, 1.09]} \\
Implied response of added reviews, $\eta$ & \multicolumn{2}{r}{0.55 [0.18, 1.15]} \\
The same ratio on the paired sample & \multicolumn{2}{r}{0.59 [0.41, 0.89]} \\
Implied $\eta$ on the paired sample & \multicolumn{2}{r}{0.38 [0.09, 0.83]} \\
\midrule
\multicolumn{3}{l}{\textit{Panel C. Sensitivity, Equation~(\ref{eq:detect}), relative to the band below}} \\
 & Model & Observed \\
Natural counts & 1.08 & 0.83 \\
Equal counts & 0.52 & 0.47 \\
\bottomrule
\end{tabular}
\par\vspace{2pt}{\footnotesize\raggedright Panel A averages each scale term over the estimation cells; regression weights and heterogeneity keep the product from matching exactly, so it is an illustration. Panel B: $a$ and $b$ use same-game flows in the band relative to the change in the band below; $w$ uses the slope of log praise and log complaints on the short-playtime score before launch; the first ratio uses the band-specific samples of Table~\ref{tab:pass} and the second the paired sample of Panel A; intervals come from 499 replicates of a bootstrap of whole game histories. Panel C: model values use the log-odds ratio relative to the band below, with mean $R(1-R)$ and median counts in each band; observed values are the ratio of correlations in Table~\ref{tab:knows} (natural counts and 20 reviews per band), relative to the band below.\par}
\end{table}

%% file: generated/tab_knows.tex
\begin{table}[H]
\centering
\caption{Sensitivity: what each stream knows, on common game-months}
\label{tab:knows}
\footnotesize
\begin{tabular}{lrrrrl}
\toprule
Measure & \multicolumn{2}{c}{3--4 hours} & \multicolumn{2}{c}{2--3 hours} & Threshold-specific \\ & Before & After & Before & After & change [95\% CI] \\
\midrule
\multicolumn{6}{l}{\textit{Panel A. Correlation with the short-playtime score, within game and month}} \\
Log odds, natural counts & 0.36 & 0.22 & 0.40 & 0.30 & $-$0.03 [$-$0.13, 0.06] \\
Log odds, natural counts, cells 40+/60+ & 0.47 & 0.34 & 0.50 & 0.41 & $-$0.04 [$-$0.16, 0.07] \\
Percent, natural counts & 0.54 & 0.24 & 0.57 & 0.30 & $-$0.04 [$-$0.14, 0.06] \\
Log odds, 20 reviews per band & 0.27 & 0.08 & 0.28 & 0.18 & $-$0.09 [$-$0.19, 0.00] \\
Log odds, 40 reviews per band & 0.38 & 0.12 & 0.40 & 0.27 & $-$0.13 [$-$0.28, $-$0.01] \\
Percent, 20 reviews per band & 0.45 & 0.11 & 0.45 & 0.23 & $-$0.11 [$-$0.22, $-$0.01] \\
\midrule
\multicolumn{6}{l}{\textit{Panel B. Rank separation of turn and calm months (AUC), log-odds change}} \\
Turn: short-playtime score falls 10+ points & 0.69 & 0.62 & 0.71 & 0.62 & +0.02 [$-$0.13, 0.17] \\
Turn: short-playtime score falls 15+ points & 0.76 & 0.80 & 0.83 & 0.54 & +0.33 [0.11, 0.54] \\
Turn: 10+ points, short-playtime 60+ reviews & 0.69 & 0.55 & 0.73 & 0.65 & $-$0.06 [$-$0.27, 0.16] \\
Turn: 10+ points, bands 40+ reviews & 0.74 & 0.62 & 0.83 & 0.77 & $-$0.06 [$-$0.24, 0.14] \\
\bottomrule
\end{tabular}
\par\vspace{2pt}{\footnotesize\raggedright Game-months where both bands meet the count rule, 1,878 before launch and 1,241 after. Equal-count rows draw the stated number of reviews per band in each month and average over 40 draws. The threshold-specific change is the change above the threshold minus the change below it, from a bootstrap of whole game histories. Panel B counts 72 and 32 turn months in its first row.\par}
\end{table}

%% file: generated/tab_display.tex
\begin{table}[H]
\centering
\caption{Display: a fixed tier against recalibrated thresholds, on common game-months}
\label{tab:display}
\footnotesize
\begin{tabular}{lrrrrl}
\toprule
Measure & \multicolumn{2}{c}{3--4 hours} & \multicolumn{2}{c}{2--3 hours} & Threshold-specific \\ & Before & After & Before & After & change [95\% CI] \\
\midrule
Fixed tier: hit rate (\%) & 42 & 12 & 47 & 41 & $-$23 [$-$50, 3] \\
Fixed tier: false alarms (\%) & 14 & 8 & 18 & 11 & +1 [$-$5, 7] \\
\midrule
Log-odds fall, thresholds at 5\% false alarms: hits (\%) & 26 & 19 & 31 & 25 & $-$2 [$-$26, 20] \\
Log-odds fall, thresholds at 10\% false alarms: hits (\%) & 32 & 28 & 38 & 28 & +6 [$-$22, 25] \\
Log-odds fall, thresholds at 20\% false alarms: hits (\%) & 46 & 38 & 49 & 44 & $-$3 [$-$32, 22] \\
Log-odds fall, at the tier's false-alarm rate: hits (\%) & 39 & 25 & 46 & 28 & +4 [$-$24, 28] \\
\midrule
Gain over the tier at the same false alarms (points) & $-$3 & 12 & $-$1 & $-$12 & +26 [0, 52] \\
\midrule
Fixed tier: sensitivity, $d'$ & 0.86 & 0.31 & 0.86 & 1.01 & $-$0.71 [$-$1.89, 0.16] \\
Fixed tier: criterion, $c$ & 0.64 & 1.25 & 0.50 & 0.74 & +0.37 [0.00, 0.93] \\
Log-odds statistic: sensitivity from AUC, $d_a$ & 0.71 & 0.44 & 0.80 & 0.43 & +0.10 [$-$0.49, 0.69] \\
\bottomrule
\end{tabular}
\par\vspace{2pt}{\footnotesize\raggedright A hit is a detection in a turn month, when the short-playtime score falls 10+ points below its trailing six-month median (72 before launch, 32 after); a false alarm is a detection in a calm month, when it stays within 5 points (545 and 479). Thresholds for the log-odds rule are set on other games within the same period and band. The last column of the gain row nets out the pre-launch gains. Tiers use the store's label boundaries at 95, 80, 70, 40 and 20 percent. $d'$ and $c$ use the log-linear correction for extreme rates; $d_a$ is $\sqrt{2}$ times the normal quantile of the AUC.\par}
\end{table}

%% file: generated/tab_public.tex
\begin{table}[H]
\centering
\caption{The reconstructed public score}
\label{tab:public}
\footnotesize\setlength{\tabcolsep}{4pt}
\begin{tabular}{lrrrrrrrr}
\toprule
Period & \multicolumn{4}{c}{Turn months} & \multicolumn{2}{c}{Calm months} & \multicolumn{2}{c}{Detection} \\ & $n$ & Short-playtime & Score & Label falls (\%) & $n$ & Label falls (\%) & $d'$ & $c$ \\
\midrule
\multicolumn{9}{l}{\textit{Panel A. Reconstructed 30-day score, by regime}} \\
Jan 2017--Feb 2019 & 100 & $-$16.3 & $-$5.8 & 33 & 392 & 9 & 0.88 & 0.88 \\
Mar--Sep 2019 & 24 & $-$17.2 & $-$10.6 & 58 & 52 & 16 & 1.15 & 0.39 \\
Before launch, all & 124 & $-$16.4 & $-$6.7 & 38 & 443 & 10 & 0.96 & 0.79 \\
Dec 2019--Jan 2021 & 40 & $-$14.0 & $-$1.3 & 9 & 367 & 3 & 0.58 & 1.59 \\
\midrule
\multicolumn{9}{l}{\textit{Panel B. Post-launch months with added reviews removed}} \\
Fixed monthly mass removed & 40 & $-$14.0 & $-$4.9 & 30 & 368 & 20 & 0.33 & 0.68 \\
Removed in proportion & 40 & $-$14.0 & $-$1.9 & 13 & 368 & 7 & 0.36 & 1.26 \\
\midrule
\multicolumn{9}{l}{\textit{Panel C. Hit rate (\%) at held-out false-alarm rates of 10\% and 20\%}} \\
 & \multicolumn{4}{c}{After launch} & \multicolumn{4}{c}{Before launch} \\
 & \multicolumn{2}{r}{10\%} & \multicolumn{2}{r}{20\%} & \multicolumn{2}{r}{10\%} & \multicolumn{2}{r}{20\%} \\
Fall in points & \multicolumn{2}{r}{27} & \multicolumn{2}{r}{39} & \multicolumn{2}{r}{38} & \multicolumn{2}{r}{56} \\
Fall in log odds & \multicolumn{2}{r}{19} & \multicolumn{2}{r}{33} & \multicolumn{2}{r}{40} & \multicolumn{2}{r}{51} \\
Complaint share relative to norm & \multicolumn{2}{r}{19} & \multicolumn{2}{r}{33} & \multicolumn{2}{r}{40} & \multicolumn{2}{r}{50} \\
Complaint count relative to norm & \multicolumn{2}{r}{12} & \multicolumn{2}{r}{20} & \multicolumn{2}{r}{23} & \multicolumn{2}{r}{36} \\
\bottomrule
\end{tabular}
\par\vspace{2pt}{\footnotesize\raggedright Rolling 30 days ending on the last day of the month, Steam purchases, all playtimes, windows with 50+ reviews. The short-playtime score uses a random half of under-2-hour reviews; that half is removed from the 30-day score and the other half is counted twice; every entry averages over 25 random halves. Changes in points against a trailing six-month calendar median within the regime. Labels use the store's boundaries and count conditions. $d'$ and $c$ are the sensitivity and criterion implied by hits and false alarms. Panel C sets thresholds on calm months in other games.\par}
\end{table}

%% file: generated/tab_failures.tex
\begin{table}[H]
\centering
\caption{Documented failures: change in positive share, net of quiet windows}
\label{tab:failures}
\footnotesize
\begin{tabular}{lrrrrrrll}
\toprule
Band & \multicolumn{3}{c}{Before launch} & \multicolumn{3}{c}{After launch} & Change [95\% CI] & Change, \\ & Windows & Points & (SE) & Windows & Points & (SE) & points & log odds \\
\midrule
Under 2 hours & 65 & 0.3 & (1.9) & 19 & 2.6 & (1.9) & +2.3 [$-$3.5, 8.4] & +0.13 (0.14) \\
1--3 hours & 57 & $-$2.0 & (1.6) & 16 & $-$1.6 & (1.3) & +0.4 [$-$3.6, 4.3] & +0.13 (0.16) \\
3--5 hours & 40 & $-$3.1 & (2.7) & 22 & $-$2.3 & (1.6) & +0.8 [$-$6.2, 6.7] & +0.10 (0.20) \\
5--30 hours & 85 & $-$1.0 & (1.4) & 24 & $-$2.0 & (1.0) & $-$1.1 [$-$4.0, 2.4] & $-$0.07 (0.13) \\
Over 30 hours & 83 & $-$1.1 & (1.2) & 22 & $-$3.5 & (1.4) & $-$2.5 [$-$6.2, 0.7] & $-$0.35 (0.21) \\
\midrule
All over 3 hours & 89 & $-$1.1 & (1.2) & 25 & $-$2.6 & (0.9) & $-$1.5 [$-$4.4, 1.4] & $-$0.08 (0.14) \\
All reviews & 90 & $-$1.0 & (1.2) & 25 & $-$2.9 & (0.8) & $-$1.8 [$-$4.5, 0.9] & $-$0.13 (0.11) \\
\bottomrule
\end{tabular}
\par\vspace{2pt}{\footnotesize\raggedright Change from 42--7 days before an acknowledged failure to the window from 7 days before to 14 days after, minus the same change in quiet windows in the same games (3,607 and 1,454). Windows need 20+ reviews in the band on each side, and each window has equal weight. Standard errors and intervals from a game bootstrap. 109 usable episodes in 55 games before launch (41 major), 25 in 18 after (7 major). AUC of failure against quiet windows, log-odds change, all reviews: 0.53 before, 0.64 after.\par}
\end{table}

%% file: generated/tab_updates.tex
\begin{table}[H]
\centering
\caption{Dated developer updates, in the bands on either side of the threshold}
\label{tab:updates}
\small
\begin{tabular}{lrrrrr}
\toprule
Windows & Count & 1--3 h & 3--5 h & \multicolumn{2}{c}{Gap (3--5 h minus 1--3 h)} \\
\midrule
Update windows, before launch & 2,267 & $-$0.062 & $-$0.086 & $-$0.023 & (0.018) \\
Update windows, after launch & 696 & $-$0.167 & $-$0.082 & +0.085 & (0.043) \\
Change across launch &&&& +0.108 & (0.045) \\
\midrule
Quiet windows, before launch & 16,647 & $-$0.046 & $-$0.062 & $-$0.016 & (0.008) \\
Quiet windows, after launch & 6,840 & $-$0.026 & $-$0.009 & +0.017 & (0.009) \\
Change across launch &&&& +0.033 & (0.012) \\
\midrule
Update change minus quiet change &&&& +0.075 & (0.049) \\
Same, isolated updates (1,262 and 437) &&&& +0.105 & (0.079) \\
\bottomrule
\end{tabular}
\par\vspace{2pt}{\footnotesize\raggedright Pooled Mantel--Haenszel log odds of a positive review in the 14 days after an update against the 28 days before it. All standard errors come from one bootstrap of games over update and quiet windows in both periods. Quiet windows are dates in the same games with no update in the same span. Isolated updates have no other update inside the window.\par}
\end{table}

%% file: generated/tab_thresholds.tex
\begin{table}[H]
\centering
\caption{The model at all three prompts}
\label{tab:thresholds}
\footnotesize
\begin{tabular}{llll}
\toprule
Measure & 3 hours (calibrated) & 10 hours & 30 hours \\
\midrule
Added share of praise, $a$ & 0.81 & 0.23 & 0.32 \\
Added share of complaints, $b$ & 0.57 & 0.10 & 0.15 \\
Game-months, before and after & 1,878 and 1,241 & 1,384 and 1,430 & 1,102 and 1,114 \\
\midrule
\multicolumn{4}{l}{\textit{Change in the pass-through gap, percent scale}} \\
\quad Model & $-$0.36 & +0.02 & $-$0.01 \\
\quad Observed [95\% CI] & $-$0.40 [$-$0.61, $-$0.21] & +0.01 [$-$0.11, 0.11] & $-$0.03 [$-$0.17, 0.12] \\
\multicolumn{4}{l}{\textit{Change in the pass-through gap, log-odds scale}} \\
\quad Model & $-$0.26 & +0.03 & $-$0.07 \\
\quad Observed [95\% CI] & $-$0.35 [$-$0.66, $-$0.04] & +0.21 [$-$0.14, 0.54] & $-$0.14 [$-$0.55, 0.23] \\
\multicolumn{4}{l}{\textit{Threshold-specific change in sensitivity, natural counts}} \\
\quad Model & +0.03 & +0.03 & $-$0.02 \\
\quad Observed [95\% CI] & $-$0.03 [$-$0.15, 0.07] & +0.09 [$-$0.03, 0.19] & $-$0.06 [$-$0.16, 0.05] \\
\multicolumn{4}{l}{\textit{Threshold-specific change in sensitivity, 20 reviews per band}} \\
\quad Model & $-$0.10 & +0.00 & $-$0.03 \\
\quad Observed [95\% CI] & $-$0.09 [$-$0.19, 0.00] & +0.06 [$-$0.07, 0.15] & $-$0.05 [$-$0.14, 0.08] \\
\bottomrule
\end{tabular}
\par\vspace{2pt}{\footnotesize\raggedright Each column compares adjacent bands on either side of the prompt: one hour wide at 3 and 10 hours, two hours wide at 30 hours. The response of added reviews and the part of a change carried by praise are fixed at their 3-hour values, so the 10-hour and 30-hour columns are out of sample. Model values use each band's flows, scores and counts, with its pre-launch value and the change in the band below; these values were recorded before the outcomes were computed, apart from the 10-hour percent gap, which appears among the placebo cutoffs. Intervals come from 499 replicates of a bootstrap of whole game histories, so the 3-hour intervals differ slightly from Tables~\ref{tab:gap} and~\ref{tab:knows}, which use 999. Sensitivity is the correlation with the short-playtime score within game and month.\par}
\end{table}

%% file: generated/tab_mech.tex
\begin{table}[H]
\centering
\caption{Added reviews in each game's worst months, and their word use}
\label{tab:mech}
\footnotesize
\begin{tabular}{lrrr}
\toprule
Measure & & & \\
\midrule
\multicolumn{4}{l}{\textit{Panel A. After launch, each game's worst fifth of months against its other months}} \\
 & Worst months & Other months & Gap (pp) \\
2--3 hours, positive (\%) & 81.4 & 88.1 & $-$6.7 (1.2) \\
Added reviews, 3--4 hours against 2--3 hours, positive (\%) & 96.1 & 97.0 & $-$0.9 (0.3) \\
Added reviews, $\pm$30 minutes, positive (\%) & 95.5 & 96.5 & $-$1.1 (0.3) \\
Before launch: 2--3 hours, positive (\%) & 64.7 & 82.2 & $-$17.5 \\
Before launch: 3--4 hours, positive (\%) & 71.6 & 85.6 & $-$14.0 \\
\midrule
\multicolumn{4}{l}{\textit{Panel B. Update, developer or breakage words at the threshold, English reviews, after launch (\%)}} \\
 & $\pm$30 min: below & Added & Share of added \\
Under 50 characters & 0.5 & 0.3 & 71 \\
50--199 characters & 4.0 & 3.2 & 23 \\
200+ characters & 18.7 & 11.6 & 6 \\
All lengths & 5.8 & 1.6 & \\
Added, at the below-threshold rate in each length bin & & 2.4 & \\
\bottomrule
\end{tabular}
\par\vspace{2pt}{\footnotesize\raggedright Worst months use each game's own 20th percentile of the short-playtime score; games need 5+ eligible months (153 games, 387 worst months). Standard errors from a game bootstrap. Added reviews are residuals from the band above minus the band below. The word rule covers update, patch, hotfix, nerf, rework, anymore, no longer, used to, ruin, broke, broken, dev, developer, since the, after the. It counts mentions and has low precision for reports of change.\par}
\end{table}

%% file: generated/tab_robust.tex
\begin{table}[H]
\centering
\caption{Robustness of the pass-through gap}
\label{tab:robust}
\scriptsize\setlength{\tabcolsep}{3pt}
\begin{tabular}{lrrllrrr}
\toprule
Sample or specification & Before & After & Change [95\% CI] & Log-odds change [95\% CI] & $N$ before & $N$ after & Games after \\
\midrule
Baseline & $-$0.04 & $-$0.44 & $-$0.40 [$-$0.57, $-$0.19] & $-$0.35 [$-$0.67, $-$0.03] & 1,878 & 1,241 & 167 \\
Excluding sale months & +0.00 & $-$0.53 & $-$0.53 [$-$0.89, $-$0.29] & $-$0.38 [$-$0.88, $-$0.04] & 1,024 & 698 & 146 \\
Steam purchases only & $-$0.03 & $-$0.47 & $-$0.45 [$-$0.69, $-$0.19] & $-$0.28 [$-$0.67, 0.04] & 1,440 & 942 & 140 \\
Never edited, one-hour tolerance & $-$0.03 & $-$0.44 & $-$0.40 [$-$0.61, $-$0.25] & $-$0.42 [$-$0.80, $-$0.14] & 1,769 & 1,179 & 162 \\
Never edited, no tolerance & $-$0.03 & $-$0.48 & $-$0.45 [$-$0.63, $-$0.27] & $-$0.49 [$-$0.78, $-$0.21] & 1,704 & 1,155 & 159 \\
English reviews only & $-$0.07 & $-$0.45 & $-$0.38 [$-$0.81, $-$0.06] & $-$0.47 [$-$0.96, 0.01] & 723 & 520 & 101 \\
Excluding free-to-play titles & $-$0.03 & $-$0.43 & $-$0.40 [$-$0.62, $-$0.20] & $-$0.31 [$-$0.63, $-$0.03] & 1,768 & 1,158 & 159 \\
Excluding the 10 most-reviewed games & $-$0.04 & $-$0.41 & $-$0.37 [$-$0.62, $-$0.21] & $-$0.32 [$-$0.69, $-$0.04] & 1,624 & 1,113 & 157 \\
Excluding early-access reviews & $-$0.03 & $-$0.40 & $-$0.37 [$-$0.58, $-$0.17] & $-$0.33 [$-$0.63, $-$0.06] & 1,654 & 1,153 & 155 \\
Games 12+ months past first review & $-$0.04 & $-$0.29 & $-$0.25 [$-$0.52, $-$0.07] & $-$0.25 [$-$0.76, 0.14] & 1,347 & 1,070 & 143 \\
Excluding games with removed periods & $-$0.04 & $-$0.37 & $-$0.33 [$-$0.51, $-$0.15] & $-$0.31 [$-$0.58, $-$0.04] & 1,809 & 1,201 & 164 \\
Pre-period from January 2018 & $-$0.05 & $-$0.44 & $-$0.39 [$-$0.56, $-$0.16] & $-$0.29 [$-$0.59, 0.06] & 1,255 & 1,241 & 167 \\
Equal windows, 14 months each side & $-$0.03 & $-$0.44 & $-$0.41 [$-$0.64, $-$0.15] & $-$0.29 [$-$0.62, 0.02] & 871 & 1,241 & 167 \\
Post-period calendar 2020 & $-$0.04 & $-$0.43 & $-$0.39 [$-$0.57, $-$0.22] & $-$0.38 [$-$0.76, $-$0.10] & 1,878 & 1,039 & 160 \\
Six months on each side of launch & +0.06 & $-$0.57 & $-$0.63 [$-$1.14, $-$0.19] & $-$0.26 [$-$0.79, 0.41] & 435 & 489 & 135 \\
Post-period Dec 2019--Feb 2020 & $-$0.04 & $-$0.42 & $-$0.38 [$-$4.49, 0.62] & $-$0.20 [$-$0.96, 0.61] & 1,878 & 247 & 117 \\
Games observed in both periods & $-$0.05 & $-$0.31 & $-$0.26 [$-$0.51, $-$0.10] & $-$0.17 [$-$0.62, 0.11] & 1,667 & 1,090 & 143 \\
Both periods, each game weighted equally & $-$0.03 & $-$0.39 & $-$0.36 [$-$0.70, $-$0.11] & $-$0.21 [$-$0.80, 0.25] & 1,667 & 1,090 & 143 \\
30+ per band, 50+ short-playtime & +0.02 & $-$0.47 & $-$0.48 [$-$0.68, $-$0.28] & $-$0.43 [$-$0.71, $-$0.12] & 1,348 & 885 & 137 \\
50+ per band, 100+ short-playtime & $-$0.03 & $-$0.42 & $-$0.39 [$-$0.61, $-$0.15] & $-$0.31 [$-$0.63, $-$0.02] & 799 & 526 & 100 \\
Bands of 45 minutes & +0.03 & $-$0.48 & $-$0.51 [$-$0.73, $-$0.33] & $-$0.43 [$-$0.78, $-$0.10] & 1,481 & 973 & 143 \\
Bands of 30 minutes & +0.03 & $-$0.54 & $-$0.57 [$-$0.87, $-$0.37] & $-$0.41 [$-$0.82, $-$0.07] & 1,347 & 894 & 139 \\
Short-playtime group: under 1 hour & $-$0.02 & $-$0.23 & $-$0.21 [$-$0.37, $-$0.05] & $-$0.18 [$-$0.52, 0.09] & 1,825 & 1,183 & 165 \\
Short-playtime group: English Steam purchases & $-$0.07 & $-$0.35 & $-$0.28 [$-$0.74, 0.13] & $-$0.24 [$-$0.89, 0.28] & 1,440 & 851 & 130 \\
\bottomrule
\end{tabular}
\par\vspace{2pt}{\footnotesize\raggedright Each row re-estimates the 3--4 hour minus 2--3 hour pair of Table~\ref{tab:gap}, with the rest of the specification unchanged and intervals from 299 replicates of the game-history bootstrap, so the baseline row differs slightly from Table~\ref{tab:gap}, which uses 999. Count rules are 20+ per band and 30+ short-playtime reviews unless the row says otherwise. Never-edited rows use the review's update time; the first allows an update within an hour of posting.\par}
\end{table}

%% file: generated/tab_hypotheses.tex
\begin{table}[H]
\centering
\caption{Summary of hypothesis tests}
\label{tab:hypotheses}
\footnotesize
\begin{tabularx}{\textwidth}{>{\raggedright\arraybackslash}p{0.8cm} >{\raggedright\arraybackslash}p{3.6cm} >{\raggedright\arraybackslash}X >{\raggedright\arraybackslash}p{3.3cm}}
\toprule
Test & Claim & Evidence [95\% CI] & Verdict \\
\midrule
H1 & Order holds near launch, within a small margin & Change $-$0.007 at 50 reviews per game, inside the margin; $-$0.032 at 200, outside it; lower by 2021--22 & Supported near launch at modest counts \\
H2a & Pass-through falls on the percent scale only & Log-odds gap changes $-$0.35 [$-$0.65, $-$0.04] & Not supported \\
H2b & Pass-through falls on both scales & Percent gap changes $-$0.40 [$-$0.57, $-$0.21]; log-odds interval excludes zero in 13 of 24 robustness rows & Supported; firm on percent, weaker on log odds \\
H3a & The stream loses sensitivity at natural counts & Threshold-specific change $-$0.03 [$-$0.13, 0.06] & Not supported; tail detection remains imprecise \\
H3b & Sensitivity falls at equal counts; volume offsets it & $-$0.09 [$-$0.19, 0.00] at 20 reviews per band; $-$0.13 [$-$0.28, $-$0.01] at 40 & Supported \\
H4 & Fixed tiers lose hits and turn conservative; new thresholds recover hits & Hits change $-$23 points [$-$50, 3]; criterion +0.37 [0.00, 0.93]; post-launch recalibration contrast +25 points [3, 45]; on the public score, added reviews raise the criterion by +0.32 to +0.91 & Supported on the criterion and on recalibration; the hit contrast is imprecise; on the public score, no rule beats the label at its false-alarm rate \\
H5 & Changes through complaints stay visible & Net fall in the all-review score $-$1.0 before, $-$2.9 after; change $-$1.8 [$-$4.5, 0.9] & Supported; few post-launch events \\
\bottomrule
\end{tabularx}
\end{table}

%% file: sections/07_discussion.tex
\section{Discussion}
\label{sec:discussion}

Our results show that one prompt can leave a rating's order largely intact, shrink its movement and quiet its labels, while changing far less what the stream knows. This section considers what that pattern implies for theory and for platform design, and where we expect it to apply.

\subsection{Theoretical implications}

Our first contribution is to reconcile the three theories with which we began. They give conflicting advice because they describe different forces, and each is partly right. Selection theory is right that more reviews lower noise, but added reviews do not behave like voluntary ones. Voice theory is right that added reviews respond less, about half as much in log odds, but the stream's sensitivity does not detectably leave with them. Inflation theory is right that scores compress and that fixed displays suffer, but this is not a loss of what the stream knows. A statement about a rating's informativeness should therefore say whether it concerns the order of products, the movement of the number, the sensitivity of the stream, or the display. In our setting these differ, and the difference decides the remedy.

This reconciliation depends on extending signal detection theory to rating systems. The theory has mostly been applied to human judgment and diagnostic tests \citep{greenswets1966,swets1988}. We apply it to an information system whose evidence is a stream of user contributions and whose decision rule is a fixed display, and this shift makes a design question precise. Changing who contributes affects sensitivity through signal and noise, and a fixed output rule then converts the new distribution into a new operating point without anyone choosing it. On the reconstructed public score, the estimates show added reviews raising the label's criterion and not lowering its sensitivity, so the platform's detector became more conservative as a by-product of a change on the input side. Theorizing the rating system as an IT artifact \citep{orlikowski2001}, with separate input and output sides, is what makes this visible.

The framework also separates two forces that are often treated as one improvement. More reviews and more representative reviews are distinct, and on the task of noticing change they pull in opposite directions. Per review, the added stream knows less, while its greater volume makes up most of the difference, and the out-of-sample check fits a threshold-specific break that scales with the added share. A platform that asks for reviews pays for participation in signal per review and is repaid in lower noise, but only readers who can see past the scale collect that repayment.

Which flow carries a change sets a boundary condition on these effects. A share-based rating is a ratio of two flows, and a prompt adds to them unequally. Changes that run through complaints move the ratio even when praise has been inflated, as our documented failures do, whereas changes in short-playtime evaluations run partly through praise and fade. This condition offers a way to predict when a solicitation program will mute a signal: ask which flow the relevant change moves, and how much of that flow the program has added.

The same evidence bears on how early reviews should be read. Researchers and managers often treat them as a measure of how a product is doing. Yet short-playtime reviews barely register failures that developers acknowledge, while the score over all reviews does, and after launch players over 30 hours respond most. The two benchmarks measure different things: one is how new players are finding a game, and the other is whether something has gone wrong for those already playing. Each matters to a buyer in a different way, and a rating's fate on one does not predict its fate on the other. Early reviews are therefore not a general sensor of product change.

\subsection{Implications for platform design}

These findings carry implications for platform design, which we state as design propositions, separating those that rest on our evidence from hypotheses to test. We did not build or test a new display, so they are implications of an evaluation and not a validated artifact.

Three propositions rest directly on our evidence. First, a platform should expect fixed labels to go quiet. When a prompt lifts and compresses scores, a fixed tier fires less often, for real turns and for noise alike. A platform that values the old sensitivity to downturns has to choose it again on the new scale by accepting more false alarms. At the label's own low false-alarm rate, we find no rule that clearly beats it on the public score, and a relative complaint alert gives no gain over a plain fall in points, so we do not recommend one. Second, old thresholds should not be reused, because rules of thumb learned on unprompted scores do not transfer. A five-point fall was once common in turn months and is now rare in any month, so developers and analysts who treat the recent score as an early warning should recalibrate to the post-prompt scale, as should anyone comparing scores across such a launch. Third, platforms should evaluate displays by hits and false alarms and not by movement. A score that moves less is not necessarily a worse detector, and a label that fires less is not necessarily a broken one. A platform should therefore measure its display's operating characteristic before and after any change on the input side, using benchmarks that do not depend on the score.

Two further propositions are hypotheses for testing. A platform can record which reviews were prompted, so it could report level from all reviews and change from voluntary reviews, or weight the two. Our results suggest that this separation could help, because per-review news is higher in the voluntary stream. They do not establish it, because volume offsets dilution at the stream level and because voluntary-only displays could invite manipulation or new selection. A platform could also decide whom to ask according to the signal it needs. A usage trigger asks people who are still using a product, but other designs are available: Epic Games Store asks randomly selected players who have played for more than two hours, after a play session \citep{epic2022}, and prompting people who have recently stopped playing would ask about an experience that a usage trigger misses. These alternatives change which flows are inflated, and so which changes stay visible.

Finally, moderation and asking should be treated as separate tools, since asking is not a remedy for review bombs in our data: the cost of a surge in negative reviews fell with Valve's filter in March 2019, before the prompt, and did not fall further after launch (Section~\ref{sec:channel}).

\subsection{Where the theory applies}

Although our evidence comes from one platform, the mechanism rests on three conditions that are not specific to games: the product can change after sale, the prompt reaches people who stayed, and fixed thresholds are used to read the score. We therefore expect the same logic to hold for in-app rating prompts on app stores, post-transaction prompts on marketplaces and delivery services, satisfaction surveys behind software dashboards, and feedback buttons on AI assistants, where operators watch a rating for signs of trouble. In each setting, the theory yields a testable claim: the more a program inflates the flow that carries a change, the more it mutes that change in any display with fixed rules, even when the underlying stream keeps its sensitivity. These are hypotheses for future work, since we have no evidence from those settings. The theory also suggests where the effect should be weak: where a rating is read relative to its own recent history, where the prompt samples people at random, and where most relevant changes raise complaints.

%% file: sections/08_limitations.tex
\section{Limitations and Future Research}
\label{sec:limitations}

These implications should be read within the scope of our evidence, and its limits point to several directions for further work. The precise playtime data cover \NGames popular games chosen by the collector and end in January 2021. Between-game tests extend to the wider store, but the over-time tests do not. Later data and smaller titles would show whether the pattern persists, whether readers adapted, and whether the later fall in agreement with critics reflects the prompt, changing games, or both. Studies of other platforms with in-product prompts could test how far the theory generalizes.

Our treatment of exposure leaves a second set of open questions. We infer the trigger from outcomes and infer added reviews from a jump, so platform data marking each prompted review would allow direct comparison and would test our assumptions about voluntary flows and timing, including whether the prompt moves reviews from later playtime to earlier. The prompt also moved reviewing away from the store page. Holding the prior month's score fixed changes little, but that score is a proxy for what a reviewer saw and leaves out influence within the month, so data on what each reviewer saw could separate being asked from answering without the score in view. The launch also coincides with other changes to the client and precedes the pandemic. Placebo cutoffs and dates test the specificity of the break, but they cannot remove every launch-related process, and the post-launch display outcomes all fall in 2020 or later.

The theory has been tested less severely than a preregistered design would allow. We developed the detection framework with the three-hour results in view. The prompts at 10 and 30 hours provide an out-of-sample check, but their intervals are wide and they come from the same platform and period, so a test on another platform, set out in advance, would be stronger.

Several measures could be improved. Short-playtime reviewers are a selected group, our sample restrictions narrow that selection without removing it, the reliability of their score falls after launch, and the score does not register documented failures. Our failure sample rests on a single coding pass, blind to review data, from the first part of each announcement, and so does the change-word check. The failure sample has few post-launch episodes, and it reflects what developers chose to acknowledge. A second, independent coding and a larger set of independently documented events would allow a sharper test, including separate estimates for technical failures and contested decisions. The interval at natural counts admits a threshold-specific fall of up to \CorrNatLoAbs, about a third of the pre-launch correlation, and estimates in larger cells leave room for a real loss, so a richer dataset could yet find a threshold-specific fall. Critic scores benchmark a persistent ordering and not current quality. The change-word rule has low precision for actual reports of change, and a validated text measure could test the content of added reviews more directly. The public score is a reconstruction based on a snapshot and on label rules observed in 2026, which we assume also held during the study period.

Finally, we measure what a rating knows and shows, not what buyers do with it. We do not observe purchases, refunds or later satisfaction, so we cannot put a welfare number on the quieter label or on the gains from more reviews. A higher displayed score may also shift expectations and so the short-playtime score itself, a spillover that a within-month comparison cannot isolate. An experiment that randomizes the display of level and change, or that randomizes which players are prompted and when, could trace the consequences for purchases and for developers' responses, and could separate the effect of asking on who answers from its effect on what a given person says.

%% file: sections/09_conclusion.tex
\section{Conclusion}
\label{sec:conclusion}

Platforms need to know whether asking for reviews helps or harms what a rating can tell its users, and existing theory gives conflicting answers. Separating what a rating knows from what it shows resolves much of the conflict. On Steam, one small prompt multiplied reviews, and the rating's agreement with critics changed little around launch. The score's movement changed far more: just above the prompt's threshold, scores followed short-playtime reviews far less closely, and fixed labels went quiet.

Much of that loss lies in the display. Each added review says less about change than a voluntary one, but there are many more of them, and at natural counts the stream's ability to track short-playtime evaluations has not detectably fallen relative to the stream below the threshold. When a failure raises complaints, the score still falls. The lesson is not that platforms should stop asking. A prompt changes the scale on which a rating must be read, so the rules that turn reviews into a number and a label need to change with it.

%% file: sections/10_appendix.tex
\section{Proofs}
\label{app:model}

\textbf{Proposition 1.} Let $x_g = \ln(L_g/V_g)$ and $R_g = e^{x_g}/(1+e^{x_g})$. If the prompt multiplies the odds by a common $\kappa > 1$, then $x'_g = x_g + \ln\kappa$. Because $R$ is strictly increasing in $x$, $x_g > x_h$ implies $R'_g > R'_h$, and the order of products is unchanged. For the spread, take two products with $x_g > x_h$. Then $R_g - R_h = \int_{x_h}^{x_g} R(u)(1-R(u))\, du$, and $R'_g - R'_h = \int_{x_h}^{x_g} R(u + \ln\kappa)(1 - R(u + \ln\kappa))\, du$. The integrand $R(1-R)$ is decreasing in $u$ for $u > 0$, so the difference shrinks for every pair of products whose scores are above one half. \qed

\textbf{Proposition 2.} After the prompt, $L' = L + A$ and $V' = V + B$, so $x' = \ln(L+A) - \ln(V+B)$. Differentiating, with $a = A/(L+A)$ and $b = B/(V+B)$, gives
\[
dx' = (1-a)\, d\ln L + a\, d\ln A - (1-b)\, d\ln V - b\, d\ln B.
\]
With $d\ln A = \eta\, d\ln L$ and $d\ln B = \eta\, d\ln V$, this becomes $dx' = [1 - a(1-\eta)]\, d\ln L - [1 - b(1-\eta)]\, d\ln V$. Without the prompt, $dx = d\ln L - d\ln V$. Substituting $d\ln L = -\ell$ and $d\ln V = v$, with $\ell,v \geq 0$ not both zero, and $w = \ell/(\ell+v)$, gives $dx'/dx = \phi$ in Equation~\eqref{eq:phi}, a weighted average of $1-a(1-\eta)$ and $1-b(1-\eta)$. This yields the stated bounds. By the identity in Equation~\eqref{eq:ratio}, $dR' = R'(1-R')\,dx'$ and $dR = R(1-R)\,dx$. Dividing,
\[
\frac{dR'}{dR} = \underbrace{\frac{R'(1-R')}{R(1-R)}}_{\text{ceiling factor}} \times \underbrace{\frac{dx'}{dx}}_{\text{sensor factor}}.
\]
The weighted-average form needs the sign restriction. If an event raises both flows, then $\ell < 0$, the weights leave the unit interval, and $\phi$ can lie outside the stated bounds. \qed

\textbf{Proposition 3.} With $n$ independent reviews and share $R$, the delta method gives a sampling variance of the log odds of approximately $1/[nR(1-R)]$. The signal in log odds is $dx$ before the prompt and $\phi\, dx$ after it. The ratio of signal to noise is $D = |dx|\sqrt{nR(1-R)}$ before and $D' = \phi |dx| \sqrt{n'R'(1-R')}$ after, which gives Equation~\eqref{eq:detect}. At $n'=n$, the factor under the root is below one whenever $R' > R \geq 1/2$, because $R(1-R)$ falls above one half. \qed

\section{The Repeated-Measurement Estimator}
\label{app:iv}

Let $s^{*}_{gm}$ be the latent positive share of short-playtime reviews in game $g$ and month $m$, and let the outcome satisfy $y_{gm} = \mu_g + \lambda_m + \pi s^{*}_{gm} + \epsilon_{gm}$. Divide the $n_{gm}$ reviews into two random halves with shares $s^{a}_{gm}$ and $s^{b}_{gm}$. If reviews are independent draws given $s^{*}_{gm}$, then $s^{j}_{gm} = s^{*}_{gm} + e^{j}_{gm}$ with $e^{a}$ and $e^{b}$ uncorrelated with each other and with $\epsilon$, because the outcome band shares no reviews with the short-playtime group. After removing game and month effects, $\mathrm{Cov}(y, s^{b}) = \pi\,\mathrm{Var}(s^{*})$ and $\mathrm{Cov}(s^{a}, s^{b}) = \mathrm{Var}(s^{*})$, so the ratio recovers $\pi$. This is measurement-error correction with duplicate measures \citep{gillen2019}. It differs from the split-sample design of \citet{angrist1995}, which estimates a first stage in one half of the sample and applies it to the other. Both halves share any selection into short-playtime reviewing and any influence on the game-month, so the correction removes sampling error only. On the log-odds scale each half is $\ln[(s+0.5)/(n-s+0.5)]$, which introduces nonclassical error in small cells; Appendix~\ref{app:sim} reports its size.

\section{Finite-Sample Behavior on the Log-Odds Scale}
\label{app:sim}

We simulate binomial counts in each period's actual cells. Levels are each game's observed mean log odds. The latent within-game movement has a standard deviation of 0.2 or 0.4 and the outcome moves $\pi$ times the latent, for $\pi$ of 0.5, 0.85 and 1. Over 200 replications, the mean estimate differs from the truth by at most \SimBias in absolute value, in both periods. Cells with no negative review are \BoundPre\% of the 3--4 hour cells before launch and \BoundPost\% after, and cells with at most two are \BoundTwoPre\% and \BoundTwoPost\%. The continuity correction therefore does not create a large difference between periods in this design.

\section{Coding of Documented Failures}
\label{app:coding}

The rule selects a developer announcement if its title or opening text mentions an apology, an outage, servers being down, a service disruption or interruption, connection, login or matchmaking problems, a rollback or reversion, game-breaking or critical bugs, damaged saves, lost progress, data loss, compensation, or an emergency patch. Of \FailCand candidates, \FailFTwo were coded as major failures, \FailFOne as limited failures, and \FailN as neither.

A \emph{major failure} acknowledges a failure that reached players generally: the game or its service unavailable or unplayable for many players, widespread lost progress or damaged saves, a release withdrawn or rolled back because of problems, compensation for a failure, or an explicit apology for a significant problem or decision. A \emph{limited failure} acknowledges a real problem of narrower reach. Routine patch notes, tuning reversals, scheduled maintenance, apologies for delays, and lists of known issues with no sign of serious harm were coded as neither. Doubtful cases went to the lower code. Episodes are \FailTech technical and \FailDec decisions the developer apologized for or reversed. Coding used the first 90 characters of the title and 230 of the text, in shuffled order, with no review data.

\section{Rules for Dated Updates}
\label{app:rules}

A news item counts as a developer update if it comes from the developer announcement feed; if its title contains \emph{update, patch, version, season, expansion, DLC, overhaul, rework, balance, changelog} or \emph{release notes}, or it carries Steam's patch-notes tag; and if its title does not contain \emph{hotfix, minor, small, maintenance, server restart, beta, test server, experimental, PTS, preview, teaser, roadmap, diary, community, contest, sale, discount, tournament, stream, livestream, survey} or \emph{FAQ}. Events in the same game are kept only if they are at least 14 days apart. Quiet windows use dates on a 14-day grid with a random offset for each game, retained if no update falls within the same span.

\begin{figure}[H]
\centering
\includegraphics[width=\textwidth]{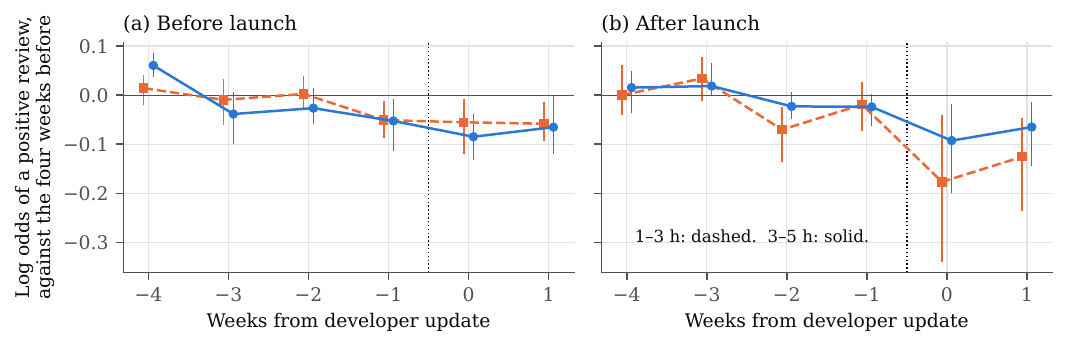}
\caption{Log odds of a positive review in each week against the four weeks before a developer update, with the windows and estimator of Table~\ref{tab:updates}. The dotted line marks the update. Bars show 95\% bootstrap intervals.}
\label{fig:updates}
\end{figure}

Figure~\ref{fig:updates} traces the update windows week by week. Before launch, both bands fall by a similar amount in the week of an update, \EvPreBelow in log odds at 1--3 hours and \EvPreAbove at 3--5 hours. After launch, the 1--3 hour band falls further than the 3--5 hour band in that week, by \EvPostBelow against \EvPostAbove, and the gap remains a week later. The 1--3 hour band also dips by \EvPostBelowDip two weeks before updates after launch, so the path supports the pooled estimates in Table~\ref{tab:updates} without offering a sharper test of them.

\section{Additional Tables}
\label{app:tables}

Table~\ref{tab:mixture} shows how the positive share of added reviews depends on what we assume about voluntary reviewing above the threshold. Each row subtracts a counterfactual voluntary flow from the reviews observed just above the threshold. The estimate barely moves with the bandwidth or with local linear flows, and it stays between \GridMin and \GridMax even when the voluntary flow and its positive share differ from those below the threshold by large margins. Under every assumption we test, added reviews are far more positive than the \OrgPos\% just below the threshold.

\input{generated/tab_mixture}

Table~\ref{tab:label} gives the unstandardized comparison behind the diagnostic score in Section~\ref{sec:display}. Its reference is the score of reviews under three hours and its outcome is the score of reviews above three hours bought on Steam, so the two share no reviews. In months when the reference falls 10 or more points, the outcome falls \LbTenDpA points before Valve's filter, \LbTenDpB points in the pre-launch months under the filter and \LbTenDpC after launch, and the tier falls in \LbTenTierA\%, \LbTenTierB\% and \LbTenTierC\% of those months. Falls of 15 or more points show the same order: \LbFifDpA, \LbFifDpB and \LbFifDpC points, and \LbFifTierA\%, \LbFifTierB\% and \LbFifTierC\%. In calm months the outcome barely moves, by \LbCalmDpA, \LbCalmDpB and \LbCalmDpC points, and the tier falls in \LbCalmTierA\%, \LbCalmTierB\% and \LbCalmTierC\% of them.

\input{generated/tab_label}

Table~\ref{tab:cutoffs} reports the search for thresholds. For each cutoff, it gives the count of reviews just above over the count just below, and the jump in the positive share, before and after launch. Before launch the count ratio lies between \CutBeforeLo and \CutBeforeHi at every cutoff. After launch, counts jump at 3, 10 and 30 hours and at higher milestones, and they show no such jump at the cutoffs between them. The positive share jumps most where counts jump most, at 3 and 1,000 hours.

\input{generated/tab_cutoffs}

Table~\ref{tab:episodes} lists every game with a month in each period in which reviews under three hours fell 12 or more points below their trailing median. It illustrates the pattern and does not test it: picking each game's largest fall over periods of unequal length is an extreme-value comparison, and the falls in the two periods differ. Civilization VI's score fell \CivPre points in June 2018 and \CivPost in September 2020. The score fell less after launch in \EpLess of the \EpG games. Rocket League in September 2020 is the exception, with a surge of negative reviews and a fall of \RocketPost points.

\input{generated/tab_episodes}

%% file: generated/tab_mixture.tex
\begin{table}[H]
\centering
\caption{Share positive among added reviews, under alternative counterfactual flows}
\label{tab:mixture}
\small
\begin{tabular}{lrl}
\toprule
Assumption & Positive (\%) & [95\% CI] \\
\midrule
Pooled, $\pm$10 minutes & 96.0 & [94.7, 97.1] \\
Pooled, $\pm$20 minutes & 96.2 & [94.7, 97.2] \\
Pooled, $\pm$30 minutes & 96.4 & [95.0, 97.4] \\
Pooled, $\pm$60 minutes & 96.9 & [95.6, 97.9] \\
Local linear flows, $\pm$30 minutes & 95.8 & \\
Pre-launch discontinuities carried over, $\pm$30 minutes & 95.6 & [94.1, 96.6] \\
Equal weight per game-month, 2,214 months & 94.9 & [94.2, 95.6] \\
\midrule
\multicolumn{3}{l}{\textit{Voluntary flow above against below, and voluntary share above against below}} \\
Flow 30\% lower, share 2 points higher & 95.1 & \\
Flow 30\% higher, share 2 points higher & 96.6 & \\
Flow 50\% lower, share 6 points higher & 94.2 & \\
Flow 50\% higher, share 6 points higher & 95.0 & \\
Flow 50\% higher, share 2 points lower & 99.4 & \\
\bottomrule
\end{tabular}
\par\vspace{2pt}{\footnotesize\raggedright Each row computes the positive share of reviews above the threshold net of a counterfactual voluntary flow, after launch. Intervals from a game bootstrap.\par}
\end{table}

%% file: generated/tab_label.tex
\begin{table}[H]
\centering
\caption{The diagnostic score over three hours}
\label{tab:label}
\footnotesize
\begin{tabular}{lrrrr}
\toprule
Period & Game-months & Under-3h change & Over-3h score change & Tier falls (\%) \\
\midrule
\multicolumn{5}{l}{\textit{Under-3h fall of 10+ points}} \\
\quad Jan 2017--Feb 2019 & 214 & $-$15.7 & $-$5.2 (1.1) & 29 \\
\quad Mar--Sep 2019 & 51 & $-$17.9 & $-$7.2 (1.7) & 43 \\
\quad Dec 2019--Jan 2021 & 85 & $-$13.8 & $-$0.8 (0.3) & 8 \\
\multicolumn{5}{l}{\textit{Under-3h fall of 15+ points}} \\
\quad Jan 2017--Feb 2019 & 85 & $-$21.2 & $-$8.4 (2.3) & 34 \\
\quad Mar--Sep 2019 & 27 & $-$22.5 & $-$10.1 (3.0) & 44 \\
\quad Dec 2019--Jan 2021 & 23 & $-$18.3 & $-$2.1 (1.0) & 17 \\
\multicolumn{5}{l}{\textit{Under-3h change within 5 points (calm)}} \\
\quad Jan 2017--Feb 2019 & 1,079 & $-$0.2 & $-$0.4 (0.1) & 13 \\
\quad Mar--Sep 2019 & 127 & 0.1 & $-$0.6 (0.3) & 10 \\
\quad Dec 2019--Jan 2021 & 888 & 0.2 & $-$0.2 (0.1) & 3 \\
\bottomrule
\end{tabular}
\par\vspace{2pt}{\footnotesize\raggedright Changes in points against a trailing median of six calendar months, computed within the regime and needing three prior months, so outcome months run Apr 2017--Feb 2019, Jun 2019--Sep 2019 and Mar 2020--Jan 2021. The over-3h score uses Steam purchases only and shares no reviews with the under-3h group. Tiers use percentage boundaries at 95, 80, 70, 40 and 20.\par}
\end{table}

%% file: generated/tab_cutoffs.tex
\begin{table}[H]
\centering
\caption{Where the review count and the share positive jump}
\label{tab:cutoffs}
\footnotesize
\begin{tabular}{rrrrrr}
\toprule
Hours & Count ratio before & Count ratio after & Jump before (pp) & Jump after (pp) & Added reviews positive (\%) \\
\midrule
1 & 0.91 & 1.02 & +0.4 (0.3) & +1.2 (0.5) & -- \\
2 & 0.93 & 0.91 & +1.0 (0.4) & +1.5 (0.5) & -- \\
3 & 0.98 & 4.50 & +0.8 (0.3) & +7.4 (0.7) & 96 \\
4 & 0.98 & 0.92 & +0.4 (0.3) & +0.1 (0.1) & -- \\
5 & 0.98 & 0.95 & +0.2 (0.3) & +0.2 (0.1) & -- \\
6 & 0.97 & 0.94 & $-$0.5 (0.3) & $-$0.1 (0.2) & -- \\
8 & 0.98 & 0.96 & +0.0 (0.3) & +0.2 (0.2) & -- \\
10 & 0.99 & 1.33 & +0.7 (0.4) & +0.7 (0.3) & 97 \\
12 & 0.98 & 0.98 & $-$0.3 (0.3) & +0.0 (0.2) & -- \\
15 & 0.99 & 0.98 & $-$0.2 (0.3) & $-$0.2 (0.2) & -- \\
20 & 0.99 & 1.04 & +0.0 (0.3) & +0.2 (0.2) & -- \\
25 & 0.98 & 0.98 & +0.2 (0.3) & $-$0.2 (0.2) & -- \\
30 & 0.99 & 1.52 & $-$0.1 (0.3) & +2.0 (0.4) & 98 \\
40 & 0.98 & 1.05 & +0.1 (0.3) & +0.7 (0.2) & -- \\
50 & 1.00 & 1.11 & +0.4 (0.3) & +0.4 (0.2) & 97 \\
100 & 0.98 & 1.18 & +1.0 (0.3) & +0.8 (0.3) & 97 \\
200 & 0.98 & 1.08 & +0.2 (0.4) & +0.5 (0.3) & -- \\
300 & 0.99 & 1.09 & $-$0.2 (0.3) & $-$0.3 (0.4) & -- \\
1000 & 1.07 & 4.09 & +1.7 (0.6) & +7.7 (1.6) & 88 \\
\bottomrule
\end{tabular}
\par\vspace{2pt}{\footnotesize\raggedright Count ratio: reviews just above over just below the cutoff. Windows of 20 minutes up to 10 hours, then 2 percent of the cutoff. Jumps are local linear with game fixed effects; bandwidth 30 minutes up to 6 hours, then 5 percent of the cutoff. The added-review share is shown where the count ratio exceeds 1.1. The 3-hour ratio here uses $\pm$20 minutes, so it differs from the $\pm$30-minute ratio in the text.\par}
\end{table}

%% file: generated/tab_episodes.tex
\begin{table}[H]
\centering
\caption{Same game, two episodes (illustration)}
\label{tab:episodes}
\small
\begin{tabular}{lrrrrrr}
\toprule
Game & \multicolumn{3}{c}{Before launch} & \multicolumn{3}{c}{After launch} \\ & Month & Under 3 h & Score & Month & Under 3 h & Score \\
\midrule
Civilization VI & 2018-06 & $-$15.4 & $-$23.6 & 2020-09 & $-$14.3 & $-$1.3 \\
MORDHAU & 2019-08 & $-$17.5 & $-$23.4 & 2021-01 & $-$12.7 & $-$3.9 \\
Assassin's Creed Origins & 2018-09 & $-$19.3 & $-$16.1 & 2020-06 & $-$12.0 & $-$3.6 \\
No Man's Sky & 2019-08 & $-$16.3 & $-$15.3 & 2020-04 & $-$15.1 & $-$3.0 \\
Rocket League & 2017-04 & $-$42.4 & $-$14.6 & 2020-09 & $-$20.0 & $-$24.1 \\
Monster Hunter: World & 2019-08 & $-$18.2 & $-$9.3 & 2020-03 & $-$12.2 & $-$1.5 \\
Dying Light & 2018-07 & $-$12.7 & $-$2.2 & 2020-05 & $-$13.0 & $-$0.4 \\
\midrule
Mean && $-$20.3 & $-$14.9 && $-$14.2 & $-$5.4 \\
\bottomrule
\end{tabular}
\par\vspace{2pt}{\footnotesize\raggedright All games with a month in each period in which the under-3-hour score fell 12+ points below its trailing six-month median, with 60+ under-3-hour reviews and 300+ reviews in that month. For each game and period the month with the largest under-3-hour fall is shown. Changes in points; the score uses all reviews.\par}
\end{table}